\documentclass[a4paper,11pt]{article}
\usepackage{graphicx,amssymb,amstext,amsmath,mathabx}
\usepackage{color, yfonts, tcolorbox}
\usepackage{floatflt}

\usepackage{float}
\usepackage{esint}
\usepackage{subcaption}
\usepackage{cite}
\usepackage{mathtools}
\usepackage{commath}
\usepackage{hyperref}

\definecolor{cbl}{rgb}{0,0,1}                

\newcommand{\bc}{\begin{center}}
\newcommand{\ec}{\end{center}}
\def\ba#1{\begin{array}{#1}\displaystyle}
\newcommand{\ea}{\end{array}}

\newcommand{\beq}{\begin{equation}}
\newcommand{\eeq}{\end{equation}}
\newcommand{\beqa}{\begin{eqnarray}}
\newcommand{\eeqa}{\end{eqnarray}}

\newcommand{\bi}{\begin{itemize}}
\newcommand{\ei}{\end{itemize}}

\newcommand{\bra}{\langle}
\newcommand{\ket}{\rangle}

\newcommand{\TT}{{\cal T}}

\DeclareMathOperator{\sgn}{sgn}

\def \be {\begin{equation}} 
\def \ee {\end{equation}}

\begin{document}
\begin{titlepage}
\vspace{0.2cm}
\begin{center}

{\large{\bf{Symmetry Resolved Entanglement of Excited States in Quantum Field Theory I: Free Theories, Twist Fields and Qubits}}}

\vspace{0.8cm} 
{\large Luca Capizzi{\LARGE $^{\star}$}, Olalla A. Castro-Alvaredo$^\heartsuit$,  Cecilia De Fazio$^\clubsuit$,\\ Michele Mazzoni$^\spadesuit$, and Luc\'ia Santamar\'ia-Sanz$^\diamondsuit$}

\vspace{0.8cm}
{\small
{\LARGE $^{\star}$}  SISSA and INFN Sezione di Trieste, via Bonomea 265, 34136 Trieste, Italy\\
\medskip

$^{\heartsuit,\spadesuit}$ Department of Mathematics, City, University of London, 10 Northampton Square EC1V 0HB, UK\\
\medskip

$^{\clubsuit}$ School of Physics and Astronomy, University of Nottingham, Nottingham, NG7 2RD, UK\\
\medskip

\hspace{-0.5cm}$^\diamondsuit$ Departamento de F\'isica Te\'orica, At\'omica y \'Optica, Universidad de Valladolid, 47011 Valladolid, Spain\\
}
\end{center}

\medskip
\medskip
\medskip
\medskip

The excess entanglement resulting from exciting a finite number of quasiparticles above the ground state of a free integrable quantum field theory has been investigated quite extensively in the literature. It has been found that it takes a very simple form, depending only on the number of excitations and their statistics. There is now mounting evidence that such formulae also apply to interacting and even higher-dimensional quantum theories. In this paper we study the entanglement content of such zero-density excited states focusing on the symmetry resolved entanglement, that is on { 1+1D} quantum field theories that possess an internal symmetry. The ratio of charged moments between the excited and grounds states, from which the symmetry resolved entanglement entropy can be obtained, takes a very simple and universal form, which in addition to the number and statistics of the excitations, now depends also on the symmetry charge. Using form factor techniques, we obtain both the ratio of moments and the symmetry resolved entanglement entropies in complex free theories which possess $U(1)$ symmetry. The same formulae are found for simple qubit states.

\noindent 
\medskip
\medskip
\medskip
\medskip


\vfill
\noindent 
{\LARGE $^{\star}$} lcapizzi@sissa.it\\
{$^\heartsuit$}o.castro-alvaredo@city.ac.uk\\
$^\clubsuit$ cecilia.defazio@nottingham.ac.uk\\
$^\spadesuit$ michele.mazzoni.2@city.ac.uk\\
$^\diamondsuit$ lucia.santamaria@uva.es

\hfill \today

\end{titlepage}
\section{Introduction}
The study of entanglement measures in the context of low-dimensional quantum field theory is a very active field of research within theoretical physics. Activity has been particularly intense since the early work of Calabrese and Cardy \cite{Calabrese:2004eu} which both extended previous results \cite{CallanW94,HolzheyLW94} and also, crucially, brought those results to the attention of a much wider scientific community. These theoretical results, in conjunction with numerical and analytical work in integrable spin chain models \cite{latorre1,Latorre2,Jin,latorre3}, revealed how certain entanglement measures, i.e. the entanglement entropy \cite{bennet}, display universal scaling at conformal critical points. This observation has many implications, a very important one being that computing the entanglement entropy of a pure state is one of the most numerically effective ways of checking for criticality and, if found, determining the universality class the model belongs to.  

A recent development in this field is the growing interest in a type of entanglement termed symmetry resolved entanglement. In the context of conformal field theory (CFT), a definition of this quantity was put forward in \cite{GS}, where it was related to correlation functions of generalised (or composite) branch point twist fields. The role of symmetries and the contribution of symmetry sectors to the total entanglement was also studied in \cite{german3} simultaneously and independently of \cite{GS}. In the context of entanglement, branch point twist fields were first introduced in \cite{Calabrese:2004eu} as associated to conical singularities in conformal maps and in \cite{entropy,benext} as symmetry fields associated to cyclic permutation symmetry in 1+1D  quantum field theories (both critical and gapped). The basic idea is that in theories that possess an underlying symmetry (say $U(1)$ symmetry in a complex free boson theory or in sine-Gordon theory) entanglement can be expressed as a sum over contributions from different symmetry sectors. Remarkably such contributions are experimentally measurable \cite{expSRE1,expSRE2,expSRE3,expSRE4}, which provides further motivation to study this quantity.  In order to make these statements more transparent, let us introduce some basic notation. Let $|\Psi\ket$ be a pure state of a 1+1D quantum field theory (QFT) and let us define a bipartition of space into two complementary regions $A$ and $\bar{A}$ so that the Hilbert space of the theory $\mathcal{H}$ also decomposes into a direct product $\mathcal{H}_A \otimes \mathcal{H}_{\bar{A}}$. Then the reduced density matrix associated to subsystem $A$ is obtained by tracing out the degrees of freedom of subsystem $\bar{A}$ in
\beq
\rho_A=\mathrm{Tr}_{\bar{A}}(|\Psi\ket \bra \Psi|)\,,
\eeq
and the von Neumann and $n$th R\'enyi entropy of a subsystem $A$ are defined as
\beq
S=-\mathrm{Tr}_A(\rho_A \log \rho_A)\quad \mathrm{and} \quad S_n=\frac{\log(\mathrm{Tr}_A \rho_A^n)}{1-n}\,.
\label{SS}
\eeq
where $\mathrm{Tr}_A \rho_A^n:={\mathcal{Z}}_n/{\mathcal{Z}}_1^n$ can be interpreted as the normalised partition function of a theory constructed from $n$ non-interacting copies or replicas of the original model. As is well known, $S=\lim_{n\rightarrow 1} S_n$. 
\medskip

 In the presence of an internal symmetry, we can also define a symmetry operator $Q$ and its projection onto subsystem $A$, $Q_A$. By construction, we have that $[Q_A, \rho_A]=0$ and if $q$ is the eigenvalue of operator $Q_A$ in a particular symmetry sector, then ${\mathcal{Z}}_n(q)=\mathrm{Tr}_A(\rho_A^n \mathbb{P}(q))$ with $\mathbb{P}(q)$ the projector onto the symmetry sector of charge $q$, can be identified as the symmetry resolved partition function. In terms of this object, the symmetry resolved entanglement entropies (SREEs) can be written as
\beq
S_n(q)=\frac{1}{1-n} \log\frac{{\mathcal{Z}}_n(q)}{{\mathcal{Z}}_1^n(q)}\quad \mathrm{and} \quad S(q)=\lim_{n\rightarrow 1} S_n(q)\,.
\label{sym}
\eeq
As discussed in \cite{GS} these quantities can best be obtained in terms of their Fourier modes, the  charged moments $Z_n(\alpha)=\mathrm{Tr}_A(\rho_A^n e^{2\pi i\alpha Q_A})$ as
\beq
\mathcal{Z}_n(q)=\int_{-\frac{1}{2}}^{\frac{1}{2}} d\alpha \, Z_n(\alpha) e^{-2\pi i\alpha q}\,,
\label{moments}
\eeq
where we have assumed for convenience that we are dealing with $U(1)$ symmetry as will be mostly the case in this paper (for discrete symmetries the integral can be replaced by a sum). The factor $2\pi$ in the exponential can be absorbed into the normalisation of the parameter $\alpha$, but the present normalisation is chosen for convenience as it leads to more elegant expressions later on. 

Starting from these basic ideas, SREEs have been computed and discussed for many classes of models, ranging from
 1+1D  CFTs \cite{GS,Laflorencie_2014,german3,FG,Bonsignori_2020,Capizzi_2020,Murciano_2021,Chen_2021,capizzi2021,dubailmur,EID}, to free \cite{Bonsignori_2019,Murciano_2020,Horvath_2021}  and interacting integrable quantum field theories  \cite{FFSRE,horvath2021branch,pottsSRE}, holographic settings \cite{newHo, newHoo, Zhao_2021,weisenberger2021,znwm-22}, lattice
models \cite{Laflorencie_2014,german3,FG,Bonsignori_2019,Bonsignori_2020,Fraenkel_2020,BHM,BCM,MDGC,CCDGM,PBC,TR,MRC2}, out of equilibrium  \cite{FG, PBC, neven2021,fraenkel2021,parez2021exact,parezlatest,sh-22} and for systems with more exotic types of dynamics \cite{TRVC,KUFS,KUFS2,operate,CLSS,tophases}. 

\medskip
The aim of this paper is to study the SREEs in the context of zero-density excited states in 1+1D gapped systems in the scaling limit.  Consider a bipartition, as outlined above, in a system of total length $L$ and a subsystem of length $\ell$ such that  the quantity $r=\ell/L$ in the scaling limit $\ell, L \rightarrow \infty$ remains finite with $r\in [0,1]$. In this scaling limit, a zero-density excited state is simply a QFT state describing a finite number of excitations above the ground state. In a series of papers \cite{excited,excited2,excited3,excited4} the increase of the entanglement entropies and logarithmic negativity\footnote{In some of these papers more complex bipartitions were also considered, e.g. multiple disconnected regions.} with respect to their ground state values was computed and found to take a remarkably universal and simple form. They depend only on $r$, on the number of excitations and on their statistics. The results were originally derived by employing the branch point twist field approach in free fermion and free boson theories, thus might seem quite limited. However, it was argued in \cite{excited} (and illustrated on the example of one and two magnon states) that the formulae should hold much more generally, for interacting and even higher-dimensional theories\footnote{In \cite{excited4} the same formulae were shown to hold for free bosons in any dimension if $r$ is replaced by the ratio of generalised volumes.}, as long as a notion of localised excitations exists. These claims have been substantiated through additional recent results. In particular, a series of works by Rajabpour and collaborators \cite{ali1,ali2,ali3,ali4,ali5,ali6} has expanded previous work in various directions: by obtaining finite volume corrections, new formulae for systems where quasiparticles are not localised, and finally by establishing that the formulae indeed hold for generic magnon states, thus also in interacting theories, in \cite{ali6}. Similar formulae have also been found for interacting higher-dimensional theories in \cite{Angel_Ramelli_2021} and even in the presence of an external potential, arising from a semiclassical limit \cite{mussvit}. Indeed, the formulae found in \cite{excited} were not entirely unexpected as they can be derived for semiclassical systems \cite{excitedalba}, however their wide range of applicability, well beyond the semiclassical regime, as well as their derivation in the context of QFT were new. 

\medskip
In this paper we want to combine these two topics, symmetry resolved entropies and excited states, to investigate how the entropy of excited states may be seen as a sum over symmetry sectors in the presence of an internal symmetry. We will focus our attention on the complex free fermion and boson theories. The total excited state entanglement of (real) free fermions and bosons was studied in \cite{excited2,excited3} and their symmetry resolved entanglement in the ground state was studied in \cite{Horvath_2021}. This paper can be seen as a generalisation of these works. 

{ Our motivation to study these types of states from this viewpoint is, first and foremost, to provide exact formulae for the SREEs of at least a class of excited states in 1+1D QFT. This is interesting because the SREE of the ground state of 1+1D QFTs has generally a very complicated form, only accessible perturbatively in some parameter, as discussed in many papers \cite{GS,Laflorencie_2014,german3,FG,Bonsignori_2020,Capizzi_2020,Murciano_2021,Chen_2021,capizzi2021,dubailmur,EID,Bonsignori_2019,Murciano_2020,Horvath_2021,FFSRE,horvath2021branch,pottsSRE}. For the present states, it is possible to show that their SREEs are as complex as those of the ground state, i.e. knowing the former is sufficient to know the latter. Moreover, for special cases when the ground state is trivial, the SREEs can be obtained exactly.
Further motivation is provided by the fact that, contrary to the total entropy, the SREEs are entanglement measures that allow us to distinguish between charged and neutral excitations, even if in the present paper we focus only on charged particles. An example where both charged and neutral excitations are present is the sine-Gordon model in the interacting regime, which was studied in \cite{horvath2021branch}.}
\medskip

Our main results can be summarised as follows. 

Let $Z_n^\Psi(L,\ell,\alpha)$ be the charged moments of the symmetry resolved $n$th R\'enyi entropy of a connected region of length $\ell$, in a pure state $|\Psi\ket^n_L$ of an $n$-replica theory in finite volume $L$. Then, the ratio of moments
\beq
\lim_{L \rightarrow \infty}\frac{Z_n^\Psi(L,rL; \alpha)}{Z_n^0(L,rL; \alpha)}=:M_n^\Psi(r;\alpha)\,,
\label{eq_Mratio}
\eeq
between the state  $|\Psi\ket^n_L$ and the ground state $|0\ket^n_L$, in the infinite volume limit with $r$ fixed, is given by a universal formula, which depends very simply on $r$ and $\alpha$. There are two particularly useful cases from which more general formulae can be constructed. When $|\Psi\ket^n_L=|1^\epsilon \ket^n_L$ is a state of a single particle excitation with $U(1)$ charge $\epsilon$  we have that 
\beq
M_n^{1^\epsilon}(r;\alpha)=e^{2\pi i \epsilon \alpha} r^n + (1-r)^n\,,
\label{una}
\eeq
whereas for a state of $k$ identical excitations of charge $\epsilon$ we have that
\beq
M_n^{k^\epsilon}(r;\alpha)= \sum_{j=0}^k [f_j^k(r)]^n e^{2\pi i j \epsilon \alpha}\,,
\label{for1}
\eeq
where $ f_j^k(r):={}_kC_{j} \, r^j (1-r)^{k-j}$ and ${}_kC_{j}=\frac{k!}{j!(k-j)!}$ is the binomial coefficient. Formula (\ref{for1}) is the building block for all other results (formula (\ref{una}) is the $k=1$ case of (\ref{for1})). A generic state comprising $s$ groups of $k_i^{\epsilon_i}$ identical particles of charge $\epsilon_i$ will have 
\beq
M_n^{k_1^{\epsilon_1}\ldots k_s^{\epsilon_s}}(r;\alpha)=\prod_{i=1}^s M_n^{k_i^{\epsilon_i}}(r;\alpha)\,.
\label{general}
\eeq
For $\alpha=0$ these formulae reduce to those found in \cite{excited,excited2,excited3,excited4}. However, whereas in those papers the results represented the difference between ground state and excited state entanglement entropies, in this case they represent the ratio of charged moments, rather than the SREEs themselves. Thus, physically speaking, their interpretation is different. In order to obtain the SREE it is necessary to isolate the charged moments of the excited state. This can be easily done analytically, as we will see later and allows us to write the SREEs in terms of the ground state entropies. For instance, for the same state $|\Psi\ket^n_L=|1^\epsilon \ket^n_L$  considered above, the SREEs (R\'enyi and von Neumann) are given by
\beq
S^{1^\epsilon}_n(r;q)=\frac{1}{1-n}\log\frac{\mathcal{Z}^{1^\epsilon}_n(r,q)}{(\mathcal{Z}^{1^\epsilon}_1(r,q))^n}=
\frac{1}{1-n} \log\frac{
\mathcal{Z}_n^0(q-\epsilon) r^n +\mathcal{Z}_n^0(q)(1-r)^n}{(\mathcal{Z}_1^0(q-\epsilon) r +\mathcal{Z}_1^0(q)(1-r))^n}\,, 
\label{SnR1}
\eeq
and 
\beqa
S^{1^\epsilon}_1(r;q)&=&- \frac{\mathcal{Z}_1^0(q-\epsilon) r\log r+\mathcal{Z}_1^0(q) (1-r)\log(1-r)+[r \partial_n\mathcal{Z}_n^0(q-\epsilon)+(1-r)\partial_n\mathcal{Z}_n^0(q)]_{n=1}}{r \mathcal{Z}_1^0(q-\epsilon) + (1-r) \mathcal{Z}_1^0(q) } \nonumber\\
&&  +\log(\mathcal{Z}_1^0(q-\epsilon) r +\mathcal{Z}_1^0(q)(1-r))\,,
\label{S1R1}
\eeqa
in terms of the ground state partition functions and their derivatives, which can be related back to the ground state entropies. Here $\mathcal{Z}^{\Psi}_n(r,q)$ are the symmetry resolved partition functions in the state $|\Psi\ket$ and $\mathcal{Z}^{0}_n(q)$ are those of the ground state, which are independent of $r$ in the scaling limit considered here. The formulae for the SREEs of other states are rather cumbersome and we discuss more general cases in Section~\ref{SREEs4}. 

\medskip
This paper is organized as follows: In Section~\ref{building} we review the branch point twist field approach to entanglement measures and its application to the study of zero-density excited states. In Section~\ref{main} we describe how branch point twist fields can be employed to obtain the ratio of charged moments and present key aspects of the computation for complex free bosons and fermions. In Section \ref{SREEs4} we discuss how the SREE of excited states can be obtained from the ratio of charged moments. In Section \ref{qubits2} we recall how the same results can be obtained from simple qubit states. These have coefficients that represent the probabilities of finding a certain number of excitations in a certain space region. In this case the SREE can be obtained explicitly. We conclude in Section \ref{theend}. In Appendix A we summarize the form factor calculation for complex free bosons and in Appendix B for complex free fermions.  In Appendix C we discuss the finite-volume expansion of the ground state two-point function of composite twist fields. 

\section{Building Blocks}
\label{building}
In this section we review very briefly the definition of (composite) branch point twist fields and their role in the computation of the SREEs of excited states.

\subsection{Entropy of Excited States and Branch Point Twist Fields}
It has been known for a long time that at least in 1+1D all standard entanglement measures, such as the von Neumann and R\'enyi entropies, can be expressed either in terms of correlators of special quantum fields known as branch point twist fields (for QFT) \cite{entropy} or in terms of local operators (for spin chain models) \cite{fractal,permutation}. In both cases the fields/operators involved act on replica theories, that is models that are constructed as $n$ non-interacting copies of the original theory. The value of $n$ defines the R\'enyi index for the R\'enyi entropy, whereas the von Neumann entropy is obtained in the limit $n \rightarrow 1$. The process of replication gives rise to a new theory which is symmetric under permutation of any of the copies. This includes symmetry under cyclic permutation of copies and in QFT  this symmetry gives rise to a quantum symmetry field, the branch point twist field $\TT_n$. As it turns out, it is this cyclic permutation symmetry which plays an important role in the context of entanglement. This can be motivated by the structure of the manifold where the quantity $\mathrm{Tr}_A \rho_A^n$ takes its values, as discussed in \cite{Calabrese:2004eu,entropy} and many other places. From these considerations it emerges that the R\'enyi entropies of a connected subsystem extending from $x=0$ to $x=\ell$ can be obtained from the equal-time correlator of two branch point twist fields, thanks to the identification:
\beq
\mathrm{Tr} \rho_A^n=\varepsilon^{4\Delta_n} {}_L^n\bra \Psi|\TT_n(0)\tilde{\TT}_n(\ell)|\Psi\ket_L^n\,,
\label{trace}
\eeq
where $\varepsilon$ is a short-distance non-universal cut-off and $\Delta_n$ is the conformal dimension of the branch point twist field \cite{kniz,orbifold,Calabrese:2004eu,entropy}:
\beq
\Delta_n=\frac{c}{24}\left(n-\frac{1}{n}\right)\, \quad \mathrm{with} \quad c \,\, \mathrm{the\,\, central\,\, charge}\,.
\eeq
Note that the expression (\ref{trace}) holds both at and away from criticality, and in the latter case $\Delta_n$ is the conformal dimension of the branch point twist field in the CFT which describes the short-distance (massless) limit of the QFT under consideration. In addition, 
$\tilde{\TT}_n=\TT_n^\dagger$ is the hermitian conjugate of $\TT_n$, which from the symmetry viewpoint implements the reverse cyclic permutation of copies. Finally, as introduced earlier, $|\Psi\ket_L^n$ is a pure state in its replica version (hence the index $n$) at finite volume $L$, that is the tensor product of $n$ identical states. Because of the definitions (\ref{SS}) differences of R\'enyi or von Neumann entropies are independent of $\varepsilon$. They depend only on the ratio
\beq
R_n^\Psi(\ell,L):=\frac{{}_L^n\bra \Psi|\TT_n(0)\tilde{\TT}_n(\ell)|\Psi\ket_L^n}{{}_L^n\bra 0|\TT_n(0)\tilde{\TT}_n(\ell)|0\ket_L^n}\qquad \mathrm{with}\qquad  R_n^\Psi(r):=\lim_{L \rightarrow \infty}R_n^\Psi(rL,L)\,,
\label{ratio1}
\eeq
where $|0\ket_L^n$ is the finite-volume replica ground state. In the scaling limit this becomes a function of $|\Psi \ket^n$ and $r$ only
and, for the states considered in the Introduction, is given by the same equations (\ref{una}), (\ref{for1}) and (\ref{general}) if we set $\alpha=0$. 

Once this picture has been established, explicit computations are possible by different approaches. For instance, we may exploit conformal invariance for critical systems or employ an expansion in terms of (finite-volume) matrix elements of $\TT_n$, typically the case in integrable QFTs (see e.g. \cite{back,Isingb,FB,E8}). These matrix elements are called form factors and a programme for their computation has long been one of the great achievements of IQFT \cite{KW,SmirnovBook}. For the branch point twist field a generalised programme was developed in \cite{entropy} and thereafter applied to many different models. 

\subsection{Composite Branch Point Twist Fields}
In \cite{GS} a generalisation of the branch point twist field formulation for the symmetry resolved entanglement was proposed. The formulation is very natural and leads to the identification of the charged moments with a two-point function
\beq
Z^\Psi_n(L,\ell; \alpha)=\varepsilon^{4\Delta_n^\alpha} {}_L^n\bra \Psi|\TT_n^\alpha(0)\tilde{\TT}_n^\alpha(\ell)|\Psi\ket_L^n\,,
\label{moments2}
\eeq
in much the same spirit as (\ref{trace}). 
The new field $\TT_n^\alpha$ and its conjugate are composite twist fields (CTFs) which can be understood as the massive versions of the CFT field:
\beq
\TT^\alpha_n(y):=\,:\TT \mathcal{V}_\alpha:(y)=n^{2\Delta_\alpha-1} \lim_{x\rightarrow y} |x-y|^{2\Delta_\alpha(1-\frac{1}{n})} \sum_{j=1}^n \TT_n(y) \mathcal{V}_\alpha^j(x)\,,
\eeq
where $\mathcal{V}_\alpha$ is the symmetry field associated with the internal symmetry of the theory. For instance, for the complex free fermion and boson it will be a $U(1)$ field with $U(1)$ charge related to the index $\alpha$. $\Delta_\alpha$ is the conformal dimension of this field and $\mathcal{V}_\alpha^j$ is a copy of this field living in copy $j$ of the replica theory. In the context of entanglement, similar composite fields with the same conformal dimension
\beq
\Delta_n^\alpha:=\Delta_n+\frac{\Delta_\alpha}{n}\,,
\label{16}
\eeq
appeared first in \cite{ctheorem,Levi,BCDLR}, with the difference that, unlike in \cite{GS}, in those papers the field $\mathcal{V}_\alpha$ was not assumed to be a symmetry field but a completely generic one. 

The main result of this paper is the finding that, similar to the quantity (\ref{ratio1}), also the ratio of the moments (\ref{moments2}) between an excited state and the ground state takes a simple universal form for many theories/excited states, as reviewed in the introduction.  Thus, we are interested in the quantity
\beq
\label{eqn:moments_of_ratio}
M_n^\Psi(r;\alpha)=\lim_{L\rightarrow \infty} \frac{{}_L^n\bra \Psi|\TT_n^\alpha(0)\tilde{\TT}_n^\alpha(r L )|\Psi\ket_L^n\,}{{}_L^n\bra 0|\TT_n^\alpha(0)\tilde{\TT}_n^\alpha(rL)|0\ket_L^n\,},
\eeq
which is a function of the ratio $r$, the charge $\alpha$ and the state $|\Psi\ket^n$. Let us now discuss how these ratios may be computed in practise, employing a form factor approach.  

\section{Symmetry Resolved Entanglement of Excited States}
\label{main}

The CTF approach provides for us a natural way to obtain the ratio of two-point functions, that is the ratio of charged moments (\ref{eqn:moments_of_ratio}). Once we have computed the ratios $M_n^\Psi(r;\alpha)$ we will see that at least for free theories and for other specific types of states (such as certain qubit states) it is possible to also obtain the SREEs of the excited state. Let us start by performing our computations in complex free bosons and fermions.

\subsection{(Composite) Branch Point Twist Field Factorisation}
The key technical problem that was solved in \cite{excited2} is the question of how to evaluate finite volume matrix elements of the branch point twist field. The same question arises for the CTF. Although a finite volume form factor programme for generic local fields exists \cite{PT1,PT2} this cannot be directly employed for twist fields (its extension to this case is still an open problem). In the absence of such a programme, an alternative approach can be used for complex free theories, where the internal $U(1)$ symmetry on each replica can be exploited to diagonalise the action of the (composite) branch point twist field \cite{doubling}. In fact, this diagonalisation procedure can also be employed in infinite volume to compute the form factors of $\TT_n^\alpha$, as done in \cite{Horvath_2021}.
The idea is that we can find a factorisation
\beq \label{TTfactorizesector_boson}
\TT^\alpha_n=\prod_{p=1}^{n} \TT_{p+\alpha}\,, \quad \quad \tilde{\TT}_n^\alpha=\prod_{p=1}^{n} \TT_{-p-\alpha}\,,
\eeq 
for complex free bosons and 
\beq \label{TTfactorizesector_fermion}
\TT^\alpha_n=\prod_{p=-\frac{n-1}{2}}^{\frac{n-1}{2}} \TT_{p+\alpha}\,, \quad \quad \tilde{\TT}_n^\alpha=\prod_{p=-\frac{n-1}{2}}^{\frac{n-1}{2}} \TT_{-p-\alpha}\,,
\eeq 
for complex free fermions, of the CTFs where the factors $\TT_{p+\alpha}$ are all $U(1)$ fields with $U(1)$ charge $p+\alpha$. These $U(1)$ were employed in \cite{Horvath_2021}, albeit with a different normalisation of the parameter $\alpha$. These fields are the result of ``fusing" two $U(1)$ fields of charges $p$ and $\alpha$; the fields $\TT_p$ employed in \cite{excited2} in terms of which the branch point twist field can be decomposed, and the $U(1)$ field $\mathcal{V}_\alpha$, which for free theories has dimension 
\beqa
\Delta_\alpha &=&\frac{\alpha^2}{2}\,\qquad\qquad\,\, \mathrm{for\,\,free\,\, fermions}\,,\\
\Delta_\alpha &=&\frac{|\alpha|-\alpha^2}{2}\,\qquad \mathrm{for\,\,free\,\, bosons}\,,
\label{deltasfree}
\eeqa
The fact that in this special case both types of field are $U(1)$ fields means that their fusion is achieved just by adding their charges. 

The fields $\TT_{p+\alpha}$ satisfy the usual equal-time exchange relations for $U(1)$ fields, which involve what is termed a factor of local commutativity
 $\gamma_{p+\alpha}=\exp({2\pi i(p+\alpha)}/{n})$, that is, the phase that a neutral field $\varphi(x)$ accrues when taking a trip around the $U(1)$ field. As reviewed in \cite{excited2}, this factor is the key ingredient in determining the form factors of these fields. 

\subsection{Computation of $M_n^\Psi(r;\alpha)$}
The computation presented in \cite{excited2} for the total entanglement entropy may be easily extended to the case of the ratio  $M_n^\Psi(r;\alpha)$ in excited states. First of all, a word is due regarding the excited state $|\Psi\ket_L^n$. In general, any state in the replica QFT can be characterised in terms of the rapidities and quantum numbers of the excitations above the ground state. Considering  a free complex theory, we may define creation operators $(a_{j}^{\epsilon})^\dagger(\theta)$ where $\epsilon=\pm$ is the $U(1)$ charge of the particle, $j=1,\ldots, n$ is the copy number, and $\theta $ is its rapidity. 
Unlike the works \cite{excited,excited2,excited3,excited4} where complex theories were considered only  in order to access results for real ones, here we are interested in obtaining results for complex models. For this reason, the type of excited states that we want to consider is in fact  simpler and more natural than those studied in previous works.  The type of $k$-particle excited state that we are interested in consists of $n$ identical copies of a standard $k$-particle state
\beq
\prod_{j=1}^n(a_{j}^{\epsilon_1})^\dagger(\theta_1)(a_{j}^{\epsilon_2})^\dagger(\theta_2)\cdots (a_{j}^{\epsilon_k})^\dagger(\theta_k)|0\ket_L^n\,,
\label{sstate}
\eeq 
where $\theta_i$ are the rapidities, $j_i$ the copy numbers and $\epsilon_i=\pm$ specifies the type of complex boson/fermion that is created by the action of the creation operator $(a_{j_i}^{\epsilon_i})^\dagger(\theta_i)$. Let us start by considering the complex free boson case.

\subsection{Complex Free Boson}
In order to represent the state we need to move to a basis where the CTF action is diagonal and factorised. In this basis, the state can be expressed in terms of creation operators $\textfrak{a}_j^\dagger(\theta)$ and $\textfrak{b}_j^\dagger(\theta)$ associated with the two types of boson. They are related to the creation operators in the standard basis as \cite{excited2}
 \beq
   \textfrak{a}_p^\dagger(\theta)=\frac{1}{\sqrt{n}}\sum_{j=1}^n e^{\frac{2\pi i j p }{n}} (a_j^+)^\dagger(\theta)\qquad \mathrm{and}\qquad  \textfrak{b}_p^\dagger(\theta)=\frac{1}{\sqrt{n}}\sum_{j=1}^n e^{-\frac{2\pi i j p }{n}} (a_j^-)^\dagger(\theta)
   \label{rels}
   \eeq
where $p=1,\ldots n$.
In summary, the two sets of creation operators are simply Fourier modes of each other. This is also the case for free fermions, but the range of values of $p$ is different, in line with (\ref{TTfactorizesector_fermion}). 

As an example, let us consider the case of one single excitation $k=1$. We  will write the state as $|1^\epsilon \ket_L^n$ where $\epsilon=\pm$ represents the $U(1)$ charge of boson type. In the original basis, this would simply be the state $(a_1^\epsilon)^\dagger(\theta)(a_2^\epsilon)^\dagger(\theta)\ldots (a_n^\epsilon)^\dagger(\theta)|0\ket_L^n$, that is a state where a single complex boson of rapidity $\theta$ and charge $\epsilon$ is present in each replica. In the diagonal basis, such a state takes the form
\begin{equation}
\label{oldnotationstate}
    |1^+\ket_L^n = \sum_{\{ N^+ \}} A_n (\{ N^+ \}) \prod_{p=1}^{n}  \; [ \textfrak{a}_p^\dag (\theta) ]^{N^+_p} |0\ket_{p,L}^n\,,\quad  |1^-\ket_L^n = \sum_{\{ N^- \}} A_n (\{ N^- \}) \prod_{p=1}^{n}  \; [ \textfrak{b}_p^\dag (\theta) ]^{N^-_p} |0\ket_{p,L}^n\,,
   \end{equation}
   where the indices $\{N^\pm\}=\{N_1^+,N_1^-, \ldots, N_n^+,N_n^-\}$ are boson occupation numbers in each sector and they are constrained by the condition that they must add up to $n$
   \beq
   \sum_{p=1}^n N_p^\pm=n\,.
   \eeq
   The coefficients $A(\{N^{\pm}\})$ can be obtained systematically from the relationships (\ref{rels}) and their inverses. 
     Combining the factorisation of the CTF with the form of the state, we can then write the two-point function as
\begin{eqnarray}
\label{expansion}
&&{} _L^n\bra {1}^+ |\TT^\alpha_n (0) \Tilde{\TT}^\alpha_n (\ell) |{1}^+\ket_L^n  = \sum_{\{ N^+ \}} \,\sum_{ \{ M^+\}} A_n^* (\{ N^+ \}) \; A_n (\{ M^+ \})  \\
&&  \times \prod_{p=1}^{n} \; {}^n_{p,L}\bra {0}| [ \textfrak{a}_{p} \; (\theta) ]^{N^+_p} \;  \TT_{p+\alpha} (0) \; \TT_{-p-\alpha} (\ell)  \; [ \textfrak{a}_{p}^\dag (\theta) ]^{M^+_p} \;  |{0}\ket_{p,L}^n \;.\nonumber 
\end{eqnarray} 
\begin{eqnarray}
\label{expansion1}
&&{} _L^n\bra {1}^- |\TT^\alpha_n (0) \Tilde{\TT}^\alpha_n (\ell) |{1}^-\ket_L^n  = \sum_{\{ N^- \}} \,\sum_{ \{ M^-\}} A_n^* (\{ N^- \}) \; A_n (\{ M^- \})  \\
&&  \times \prod_{p=1}^{n} \; {}^n_{p,L}\bra {0}| [ \textfrak{b}_{p} \; (\theta) ]^{N^-_p} \;  \TT_{p+\alpha} (0) \; \TT_{-p-\alpha} (\ell)  \; [ \textfrak{b}_{p}^\dag (\theta) ]^{M^-_p} \;  |{0}\ket_{p,L}^n \;.\nonumber 
\end{eqnarray} 
This can be computed in the standard way, by inserting a sum over a complete set of states between the two $U(1)$ fields as detailed in Appendix A. A particular subtlety of this kind of computation is that, because of finite volume, the momenta/rapidities of excitations are quantised and non-zero matrix elements correspond to particular quantisation conditions that take the monodromy of the fields into account. In particular we have:
\begin{equation}
    P(\theta^\pm_i) = m \sinh\theta_i^\pm= 2\pi J^\pm_i \pm \frac{2\pi (p + \alpha)}{n} \,, \quad \quad J^\pm_i \in \mathbb{Z}\,,
    \label{quan}
\end{equation}
where $\theta^\pm_i$ are understood as rapidities of particles of type $\textfrak{a}_j^\dagger(\theta_i)$ and $\textfrak{b}_j^\dagger(\theta_i)$, respectively, which would be present in the sum over intermediate states. Similarly the rapidity $\theta$ is also quantised through $P(\theta)=2 \pi I$ for $I\in \mathbb{Z}$. 
Note the quantity $\frac{p + \alpha}{n}$ is never an integer for $\alpha \in [-\frac{1}{2}, \frac{1}{2}]$ and $p\neq n$ ($p=n$ corresponds to the identity field). This guarantees that only non-diagonal form factors (that is matrix elements involving only distinct right and left states) will be involved in the computation of the leading large-volume contribution to \eqref{expansion}. 

Once a sum over a complete set of states is inserted in  \eqref{expansion} the problem reduces to the computation of matrix elements of the $U(1)$ fields $\TT_{p+\alpha}$. Such matrix elements have been known for a long time but they were re-derived in \cite{excited2,Horvath_2021}. Because of the free nature of the theory, all matrix elements are given in terms of permanents whose basic building block are the two-particle form factors
\beq
f_{p+\alpha}^n(\theta_{12})={}_p\bra 0|\TT_{p+\alpha}(0)\,\textfrak{a}_p^\dagger(\theta_1)\textfrak{b}_p^\dagger(\theta_2)|0\ket_p=   -\tau_{p+\alpha}\sin\frac{\pi(p+\alpha)}{n}\frac{e^{\left(\frac{p+\alpha}{n}-\frac{1}{2}\right)\theta_{12}}}{\cosh\frac{\theta_{12}}{2}}\,,
\eeq
where $\tau_{p+\alpha}$ is the vacuum expectation value of $\TT_{p+\alpha}$ and $\theta_{12}=\theta_1-\theta_2$. 

In summary, all results obtained in \cite{excited2} follow through for the CTF with the replacement $p\rightarrow p+\alpha$ and the choice of an appropriate state. In particular, the ratio of the moments for an excited state of one excitation are nearly identical to formula (4.19) in \cite{excited2}, namely
\beq
M^{1^\pm}_n(r;\alpha)=\sum_{\{ N^\pm \}} |A_n (\{ N^\pm \})|^2 \prod_{p=1}^n  (N_p^\pm!) [g_{\pm(p+\alpha)}^n(r)]^{N_p^\pm}=e^{\pm 2\pi i\alpha}r^n+(1-r)^n\,,
\label{1par}
\eeq 
which is, as anticipated, the formula (\ref{una}) and where
\beq
g_p^n(r):=1-(1-e^{\frac{2\pi i p}{n}})r\,.
\label{gfun}
\eeq
For free bosons, this can be generalised to states containing $k$ identical excitations to find (\ref{for1}).
For states containing $k$ different excitations (with different rapidities and any combination of charges $\epsilon_i$) the result is
\beqa
M^{1^{\epsilon_{1}}  \dots 1^{\epsilon_k}}_n(r;\alpha)&=& \prod_{s=1}^k \sum_{\{N^\pm\}}|{C}_n(\{N^\pm\})|^2 \prod_{p=1}^n N_{p,s}^+!\,\, N_{p,s}^-! \left( g^n_{p+\alpha} (r) \right)^{ N_{p,s}^+} \left( g^n_{-p-\alpha} (r) \right)^{ N_{p,s}^-}\nonumber\\
&=&\prod_{j=1}^k\left[e^{2\pi i \epsilon_j\alpha}r^n+(1-r)^n\right]\,,
\label{re2}
\eeqa 
where each $\epsilon_j$ in the excited state is either $+$ or $-$. In particular, if all charges are identical, the product may be replaced by a power $k$. In these formulae $C_n(\{N^\pm\}$ and ${A}_n(\{N^\pm\}$ are coefficients which are determined by the form of the state in the diagonal basis.
 Both results are special cases of (\ref{general}).
These formulae are also derived  in Appendix A. 

\subsection{Complex Free Fermion}
For complex free fermions the computation is very similar, but states involving identical excitations are forbidden and the relationship between the original creation operators and those in the diagonal base is also slightly different. We now have 
 \beq
   \textfrak{a}_p^\dagger(\theta)=\frac{1}{\sqrt{n}}\sum_{j=1}^n e^{\frac{2\pi i j p }{n}} (a_j^+)^\dagger(\theta)\qquad \mathrm{and}\qquad  \textfrak{b}_p^\dagger(\theta)=\frac{1}{\sqrt{n}}\sum_{j=1}^n e^{-\frac{2\pi i j p }{n}} (a_j^-)^\dagger(\theta)
   \label{relsfermion}
   \eeq
where $p=-\frac{n-1}{2},\ldots \frac{n-1}{2}$, in terms of operators $(a_j^\pm)^\dagger(\theta)$ which anticommute for distinct values of $j$. The $U(1)$ twist field form factors are also modified to \cite{bernard}
\beq
f_{p+\alpha}^n(\theta_{12})={}_p\bra 0|\TT_{p+\alpha}(0)\,\textfrak{a}_p^\dagger(\theta_1)\textfrak{b}_p^\dagger(\theta_2)|0\ket_p=   i\tau_{p+\alpha}\sin\frac{\pi(p+\alpha)}{n}\frac{e^{\left(\frac{p+\alpha}{n}\right)\theta_{12}}}{\cosh\frac{\theta_{12}}{2}}\,.
\label{FFfer}
\eeq
The structure of a state consisting of a single particle excitation is as for the free boson, namely
\beq
|1^+\ket^n_L=\prod_{j=1}^n (a_j^+)^\dagger(\theta)|0\ket_L^n= \prod_{j=1}^n \frac{1}{\sqrt{n}} \sum_{p=-\frac{n-1}{2}}^{\frac{n-1}{2}} \omega^{jp} \textfrak{a}_p^\dagger(\theta)|0\ket_{L}^n\,,
\label{ferm1+}
\eeq 
\beq
|1^-\ket_{L}^n=\prod_{j=1}^n (a_j^-)^\dagger(\theta)|0\ket_L^n= \prod_{j=1}^n \frac{1}{\sqrt{n}} \sum_{p=-\frac{n-1}{2}}^{\frac{n-1}{2}} \omega^{-jp} \textfrak{b}_p^\dagger(\theta)|0\ket_{L}^n\,,
\label{ferm1-}
\eeq 
with $\omega=e^{-\frac{2\pi i}{n}}$. For instance, for $n=2$ we have:
\beq
|1^+\ket_{L}^2= \frac{1}{2}(i\textfrak{a}_{-\frac{1}{2}}^\dagger(\theta) -i\textfrak{a}_{\frac{1}{2}}^\dagger(\theta))(-\textfrak{a}_{-\frac{1}{2}}^\dagger(\theta) -\textfrak{a}_{\frac{1}{2}}^\dagger(\theta))|0\ket_{L}^2=-i\textfrak{a}_{-\frac{1}{2}}^\dagger(\theta)\textfrak{a}_{\frac{1}{2}}^\dagger(\theta)|0\ket_{L}^2\,,
\eeq 
and 
\beq
|1^-\ket_{L}^2= \frac{1}{2}(-i\textfrak{b}_{-\frac{1}{2}}^\dagger(\theta) +i\textfrak{b}_{\frac{1}{2}}^\dagger(\theta))(-\textfrak{b}_{-\frac{1}{2}}^\dagger(\theta) -\textfrak{b}_{\frac{1}{2}}^\dagger(\theta))|0\ket_{L}^2=i\textfrak{b}_{-\frac{1}{2}}^\dagger(\theta)\textfrak{b}_{\frac{1}{2}}^\dagger(\theta)|0\ket_{L}^2\,.
\eeq 
Similarly, for $n=3$:
\beq
|1^+\ket_{L}^3=i\textfrak{a}_{-1}^\dagger(\theta)\textfrak{a}_{0}^\dagger(\theta)\textfrak{a}_{1}^\dagger(\theta)|0\ket_{L}^3\quad,\quad |1^-\ket_{L}^3=-i\textfrak{b}_{-1}^\dagger(\theta)\textfrak{b}_{0}^\dagger(\theta)\textfrak{b}_{1}^\dagger(\theta)|0\ket_{L}^3.
\eeq
As we can see, due to the anticommutation relations amongst creation operators, many contributions now cancel each other so that the states take extremely simple forms in the new diagonal basis. 
One can easily show by induction that the general structure of the states \eqref{ferm1+} and \eqref{ferm1-} is:
\beq
|1^+\ket^n_L=e^{i\kappa}\prod_{p=-\frac{n-1}{2}}^{\frac{n-1}{2}} \textfrak{a}_p^\dagger(\theta)|0\ket_{L}^n\quad , \quad |1^-\ket^n_L=e^{-i\kappa}\prod_{p=-\frac{n-1}{2}}^{\frac{n-1}{2}} \textfrak{b}_p^\dagger(\theta)|0\ket_{L}^n
\eeq
with $\kappa$ a real parameter that can be computed for each specific state but will play not role in our computation. Making use of the factorisation \eqref{TTfactorizesector_fermion} we can expand the fermionic two-point function in terms of a sum over the form factors (\ref{FFfer}). The details are presented in Appendix B. For a state consisting of a single excitation the result is 
$$
M^{1^\pm}_n(r;\alpha)=\prod_{p=-\frac{n-1}{2}}^{\frac{n-1}{2}} g_{\pm p \pm \alpha}^n(r)=\prod_{p=-\frac{n-1}{2}}^{\frac{n-1}{2}} \left[1-(1- e^{\pm \frac{2\pi i (p+\alpha)}{n}})r \right] \,.
$$
Since the quantities $e^{\pm \frac{2\pi i p}{n}}$
are the $n$th roots of $+1$ for $n$ odd, and the $n$th roots of $-1$ for $n$ even, we can show:
\beq 
\prod_{p=-\frac{n-1}{2}}^{\frac{n-1}{2}}(x-e^{\pm\frac{2\pi i p}{n}}y)=x^n + (-y)^n\,,
\eeq
which, after setting $x=1 - r$, $y=-re^{\pm\frac{2\pi i \alpha}{n}}$ gives:
\beq
\label{eqn:g_function_identity}
\prod_{p=-\frac{n-1}{2}}^{\frac{n-1}{2}} g^n_{\pm p\pm\alpha}(r) = e^{\pm 2\pi i \alpha} r^n+ (1-r)^n\,,
\eeq
that is, the same formula as for free bosons, albeit resulting from a rather different product of $g$-functions. Similarly, all free boson formulae presented in the previous subsection are recovered for free fermions, as long as we consider only  distinct excitations. Further details are presented in Appendix B. 

\section{Symmetry Resolved Entanglement Entropies}
\label{SREEs4}
Having obtained the ratios of charged moments we now proceed to computing the SREE of excited states. To this aim, we need to isolate the charged moments of the excited state and then compute their Fourier transform as defined in (\ref{moments}). In other words, we need to multiply our results of the previous sections by the ground state correlator in the infinite-volume limit considered here. Note that this ground state correlator will generally be different for different theories, even if they all satisfy the formulae (\ref{una})-(\ref{general}). 

For (local) 1+1D QFTs, such as complex free theories, the ground state correlator in our scaling limit reduces to its disconnected part, that is the square of the vacuum expectation value (VEV) of the field $\TT_n^\alpha$.  This result follows simply from clustering of correlators in local QFT, but can also be demonstrated explicitly from the finite volume expansion of  the ground state two point function. This expansion is presented in Appendix~\ref{finiteVexp} for complex free fermions. In particular, looking at equation (\ref{eqn:finite_volume_3}) we can see how, despite the complexity of the expansion, in infinite volume the only surviving term in the sum corresponds to the product of VEVs $|\tau_{p+\alpha}|^2$. The same statement holds  for complex free bosons, where  the expansion is identical except for the permutation signs, which are absent from the formula, and a small change to the exponential factors. As mentioned earlier, it is also common to normalise the correlators by the inclusion of a UV cut-off, so that the natural quantity to compute is
\beq
Z^{\Psi}_n(r;\alpha)=Z^{0}_n(\alpha) M_n^\Psi(r;\alpha)\,\qquad {\rm with} \quad Z^{0}_n(\alpha):=\varepsilon^{4\Delta_\alpha} \bra \TT_n^\alpha \ket^2\,,
\eeq  
where $Z^{\Psi}_n(r;\alpha)$ are the charged moments of the excited state in our particular scaling limit and 
$Z^{0}_n(\alpha)$ are the moments of the ground state
where
$ \bra \TT_n^\alpha \ket$ is the VEV of the CTF. As we know, from general dimensionality arguments as can be found for instance in \cite{Zamo}, the VEV has a very particular dependence on the mass scale and the conformal dimension of the CTF. In fact, we have that 
\beq
\bra \TT_n^\alpha \ket= v_n^\alpha m^{4\Delta_n^\alpha}\,,
\eeq 
where $v_n^\alpha$ is a function that depends on the model and can be determined by requiring CFT normalisation of the CTF (that is, that the CFT two-point function has numerical coefficient of 1) and $\Delta_n^\alpha$ is given by (\ref{16}). The Fourier transform of the ground state moments has been studied in detail for free QFTs in \cite{Murciano_2020,Horvath_2021}, thus we will not revisit its computation here. Instead, we show that, assuming $Z^{0}_n(\alpha)$ to be known , it is possible to express the symmetry resolved partition functions and entropies of excited states fully in terms of those of the ground state.  The reason for this is that the functions $M_n^\Psi(r;\alpha)$ depend on $\alpha$ in an extremely simple manner, namely through factors of the form $e^{\pm 2\pi i j \alpha}$ only. Thus, in order to compute the SREE of an excited state, the only non-trivial integrals that we need to consider are of the form 
\beq
\int_{-\frac{1}{2}}^\frac{1}{2} d\alpha \, {Z}^0_n(\alpha) e^{-2\pi i \alpha (q\pm j)}=\mathcal{Z}^0_n(q\pm j)\,.
\eeq
For instance, using (\ref{una}), the simple example of a single excitation of charge $\epsilon$ gives the following relationship amongst partition functions
\beq
\mathcal{Z}^{1^\epsilon}_n(r;q)=\mathcal{Z}_n^0(q-\epsilon) r^n + \mathcal{Z}_n^0(q)(1-r)^n\,.
\eeq  
Therefore, the symmetry resolved R\'enyi and von Neumann entropies of such a state would be given by formulae (\ref{SnR1}) and (\ref{S1R1}), respectively. They can in turn be written in terms of the SREE and partition function of the ground state (i.e. eliminating derivative terms) by recalling that
\beq
\left.\partial_n \mathcal{Z}_n^0(q)\right|_{n=1}= - \mathcal{Z}_1^0(q) [S_1^0(q)-\log {Z}_1^0(q)]\,. 
\eeq
Similar relations are found for more complicated cases, such as (\ref{for1}), that is an excited state of $k$ identical excitations of charge $\epsilon$. In this case we find instead 
\beq
S^{k^\epsilon}_n(r;q)=\frac{1}{1-n}\log\frac{\sum_{j=0}^k \left[f_j^k(r)\right]^n \mathcal{Z}_n^0(q-\epsilon j)}{\left[ \sum_{j=0}^k f_j^k(r) \mathcal{Z}_1^0(q-\epsilon j)\right]^n}\,, 
\eeq
and the symmetry resolved von Neumann entropy
\beqa
S^{k^\epsilon}_1(r;q)&=&-\frac{\sum_{j=0}^k \left[ \mathcal{Z}_1^0(q-\epsilon j) f_j^k(r)\log f_j^k(r) + f_j^k(r) \left.\partial_n \mathcal{Z}_n^0(q-\epsilon j)\right|_{n=1} \right]}{\sum_{j=0}^k f_j^k(r) \mathcal{Z}_1^0(q-\epsilon j)}\nonumber\\
&& + \log \sum_{j=0}^k f_j^k(r) \mathcal{Z}_1^0(q-\epsilon j)\,.
\eeqa
Unlike for the charged moments, the entropies of other states are not simple sums/products of the formulae above. However, the kind of integrals involved are of the same type so that the computation can be performed in a similar manner for any excited state. As a last example, let us consider the ratio of charged moments for an excited state of two particles of opposite charges. We have that 
\beqa
\mathcal{Z}^{1^+1^-}_n(r;q)&=&\int_{-\frac{1}{2}}^\frac{1}{2} d\alpha \, Z_n^0(\alpha)(r^n+e^{2\pi i \alpha} (1-r)^n)(r^n+e^{-2\pi i \alpha} (1-r)^n) e^{-2\pi i \alpha q} \nonumber\\
&=&
\mathcal{Z}_n^0(q) (r^{2n}+(1-r)^{2n}) +  (\mathcal{Z}_n^0(q-1)+\mathcal{Z}_n^0(q+1))r^n (1-r)^n\,,
\eeqa  
so that the R\'enyi entropy is
\beq
S_n^{1^+1^-}(r;q)=\frac{1}{1-n} \log \frac{\mathcal{Z}_n^0(q) (r^{2n}+(1-r)^{2n}) +  (\mathcal{Z}_n^0(q-1)+\mathcal{Z}_n^0(q+1))r^n (1-r)^n}{[\mathcal{Z}_1^0(q) (r^{2}+(1-r)^{2}) +  (\mathcal{Z}_1^0(q-1)+\mathcal{Z}_1^0(q+1))r (1-r)]^n}\,,
\eeq 
from which the von Neumann entropy follows as above.
\medskip

In conclusion,  the SREE of the kind of excited states considered here can be expressed in terms of the SREE and partition function of the ground state. This statement holds for any systems where formulae (\ref{una})-(\ref{general}) apply and where the ground state contribution is well-defined, which, as we shall see below and in subsequent work \cite{partII}, includes a wide range of models, well beyond free QFTs.

{ We conclude this section by recalling that a key property of the SREEs of the ground state both in QFT \cite{GS} and interacting quantum spin chains \cite{german3} is the property of equipartition at leading order. That is, within a certain range of parameters\footnote{For massive QFT this range typically corresponds to the double limit of large subsystem size $\ell\gg 1$ and $|\log(m\epsilon)|\ll 1$ where $m$ is a typical mass scale and $\epsilon$ a UV cut-off (see e.g. \cite{horvath2021branch,pottsSRE}).} the SREEs of all charge sectors are charge independent. It is clear from the formulae above that this property also holds for the SREEs of excited states, as their charge dependence is solely encoded in the symmetry resolved partition function and entropies of the ground state. Thus if the entropy is equipartite in the ground state it will also be so in excited states. }

\section{Qubit States}
\label{qubits2}
Besides the QFT approach based on twist fields that we have presented so far, there are alternative ways in which the entanglement of excited states may be studied. In the works \cite{excited,excited2,excited3} several models and approaches were considered, including the study of the entanglement of certain qubit states. In the present context, such states are also useful as they provide a simpler way of obtaining our formulae for the ratios of charged moments, even if their associated SREEs will be different, in fact much simpler than those of QFT states. 

Considering the bipartite Hilbert space $\mathcal{H}= \mathcal{H}_A \otimes \mathcal{H}_{\bar{A}}$, where each factor can be related to the Hilbert space for $N_j$ sets of $j$ indistinguishable qubits (with $N=\sum_j N_j$), we can construct state of this Hilbert space $\mathcal{H}$ as
\begin{equation}
|\Psi_{\textrm{qb}}\ket = \sum_{\textbf{\textrm{q}} \in \prod_{j \geq 1} \{0,1,\dots,j\}^{N_j}} \sqrt{p_{\bf q}}|\textbf{\textrm{q}}\ket \otimes |\bar{\textbf{\textrm{q}}}\ket
\label{qubits}
\end{equation}
where the function $p_{\bf q}:=\prod_{i} f_{j_i}^{q_i}(r)$ (see the definition in  (\ref{for1})) represents the probability of finding a particle configuration $\textbf{\textrm{q}}=\{q_i: i=1,\ldots,N\}$ in the corresponding entanglement region and $|\bar{\textbf{\textrm{q}}}\ket$ is the state where the qubits are inverted. The entanglement entropy associated with this state captures the excess entanglement of an excited state containing $N_j$ sets of $j$ indistinguishable excitations if we assume that the probability of finding an excitation in subsystem $A$ is $r$. We identify the qubit state 1(0) with the presence (absence) of a particle and the non-trivial binomial coefficients account for the (un)distinguishability of excitations. Instead, in the case of the SREE, a similar computation as performed in \cite{excited2} gives the charged moments of the state. Notice that, as explained in \cite{GS}, only particles in subsystem $A$ acquire a phase when they go around a loop on the $n$-sheeted  Riemann surface with the Aharonov-Bohm flux inserted. This also means that, unlike for the entanglement entropies, the charged moments are not symmetric under exchange of $r$ and $1-r$, as we have seen in all our formulae so far.

Assuming that the charge operator associated with the internal symmetry is $Q=Q_A \otimes 1_{\bar{A}}+1_A \otimes Q_{\bar{A}}$, that is the charge operator can be decomposed into its projections into regions $A$ and $\bar{A}$, then 
\begin{eqnarray}
e^{2\pi i \alpha Q_A} |\Psi_{\textrm{qb}}\ket &=& e^{2\pi i \alpha Q_A} \sum_{\textbf{\textrm{q}}} \sqrt{p_{\bf q}}|\textbf{\textrm{q}}\ket \otimes |\bar{\textbf{\textrm{q}}}\ket = \sum_{\textbf{\textrm{q}}} e^{2\pi i \alpha (n^{+}_{q}-n^{-}_{q})} \sqrt{p_{\bf q}} |\textbf{\textrm{q}}\ket \otimes |\bar{\textbf{\textrm{q}}}\ket\,,
\end{eqnarray}
where the summation is over $\textbf{\textrm{q}} \in \prod_{j \geq 1} \{0,1,\dots,j\}^{N_j}$ and $n^\pm_q$ is the number of positively/ negatively charged particles in subsystem $A$ for a particular configuration ${\bf q}$. Note that $\rho_A$ and $Q_A$ share the same eigenbases and therefore commute. 
The charged moments of this qubit state are given by the usual formula
$\mathrm{Tr}_A(\rho_A^n e^{2\pi i\alpha Q_A})$ so that, adapting this formula to our state and using instead the notation $\rho_{\textrm{q}}$ and $Q_{\textrm{q}}$ to denote the reduced density matrix and charge operator associated with the qubit state (\ref{qubits}), the 
charged moments become
\begin{equation}
\label{qubitmom}
\mathrm{Tr}(\rho_\textrm{q}^n e^{2\pi i\alpha Q_\textrm{q}}) = \sum_{\textbf{\textrm{q}}^\prime} \bra \textbf{\textrm{q}}^\prime| \rho_\textrm{q}^n\,  e^{2\pi i\alpha Q_\textrm{q}}|\textbf{\textrm{q}}^\prime \ket = \sum_{\textbf{\textrm{q}}^\prime} e^{2\pi i\alpha (n^+_{\textrm{q}^\prime}-n^-_{\textrm{q}^\prime})} \bra \textbf{\textrm{q}}^\prime| \rho_\textrm{q}^n\, |\textbf{\textrm{q}}^\prime \ket \,.
\end{equation}
The density matrix can be written as:
\begin{equation}
\rho_\textrm{q} = \mathrm{Tr}_{\bar{\textrm{q}}}(|\Psi_{\textrm{qb}}\ket \bra \Psi_{\textrm{qb}}| ) = \sum_{\bar{\textbf{\textrm{q}}}^\prime} \bra \bar{\textbf{\textrm{q}}}^\prime| \Psi_{\textrm{qb}}\ket \bra \Psi_{\textrm{qb}} |\bar{\textbf{\textrm{q}}}^\prime \ket = \sum_{\bar{\textbf{\textrm{q}}}^\prime} \sum_{\textbf{\textrm{q}}} p_{\textbf{\textrm{q}}}\,  \delta_{\bar{\textrm{q}},\bar{\textrm{q}}^\prime}\,  |\textbf{\textrm{q}}\ket \bra \textbf{\textrm{q}}|\,,
\end{equation}
so that plugging this expression into (\ref{qubitmom}) we obtain
\begin{equation}
\mathrm{Tr}(\rho_\textrm{q}^n e^{2\pi i\alpha Q_\textrm{q}}) =  \sum_{\textbf{\textrm{q}}^\prime} \sum_{\bar{\textbf{\textrm{q}}}^\prime}\sum_{\textbf{\textrm{q}}} e^{2\pi i\alpha (n^+_{\textrm{q}^\prime}-n^-_{\textrm{q}^\prime})} \, p_{\textbf{\textrm{q}}}^n\,   \delta_{\bar{\textrm{q}},\bar{\textrm{q}}^\prime}\,\,  \delta_{\textrm{q},\textrm{q}^\prime}=  \sum_{\bar{\textbf{\textrm{q}}}^\prime}\sum_{\textbf{\textrm{q}}} e^{2\pi i\alpha (n^+_{\textrm{q}}-n^-_{\textrm{q}})} \, p_{\textbf{\textrm{q}}}^n\,   \delta_{\bar{\textrm{q}},\bar{\textrm{q}}^\prime}\,,
\end{equation}
which reproduces all results (\ref{una}), (\ref{for1}) and (\ref{general}) upon specifying the corresponding qubit state. For instance, for the simple case of a single excitation, the relevant state is 
\beq
|\Psi_{\textrm{qb}}\ket=\sqrt{r}|10\ket +\sqrt{1-r} |01\ket\,,
\eeq
from which (\ref{una}) is easily reproduced. 

\subsection{Symmetry Resolved Entanglement Entropies}
We close this section by noting that for qubit states, the results obtained are directly the moments of the state (i.e. we can think of the ground state as being trivial in these cases). This means that the formulae (\ref{una})-(\ref{general}) are directly the quantities we need to Fourier-transform in order to obtain the SREEs. The simplicity of the formulae allows us to obtain the SREEs exactly, something that is typically beyond reach for QFT. Noting that
\beq
\int_{-\frac{1}{2}}^{\frac{1}{2}} d\alpha \, e^{-2\pi i \alpha x}=\frac{\sin \pi x}{\pi x}=\delta_{x,0}\qquad {\rm for}\quad x\in \mathbb{Z}\,,
\label{funk}
\eeq
it is easy to show that
\beq
S^{1^\epsilon}_n(r;q)= \frac{1}{1-n} \log\left[\frac{\delta_{q,\epsilon} r^n + \delta_{q,0} (1-r)^n}{\left(\delta_{q,\epsilon} r + \delta_{q,0} (1-r)\right)^n}\right]\,,
\label{s1}
\eeq
and
\beq
S^{k^\epsilon}_n(r;q)=\frac{1}{1-n}\log\frac{\sum_{j=0}^k \left[f_j^k(r)\right]^n \delta_{q,\epsilon j}}{\left[ \sum_{j=0}^k f_j^k(r) \delta_{q,\epsilon j}\right]^n}\,,
\label{sk}
\eeq
from which the von Neumann entropies easily follow. Due to the simplicity of the states however, we can easily see that all the entropies above are identically zero, whenever any of the delta-functions is 1. This can be interpreted as the statement that the SREE does not give any additional information about these states. Another way to put this, is to say that the only property that matters in establishing formulae (\ref{s1})-(\ref{sk}) is whether particles are distinguishable or not and in both formulae  particles are identical by construction, so that specifying the charge does not add any relevant information.

The situation is different though if we consider states containing at least some distinct excitations. For instance, for a state of $k$ distinct excitations of the same charge $\epsilon$ we have that the charged moments are given by 
\beq
(r^n+e^{2\pi i \epsilon \alpha}(1-r)^n)^k = \sum_{j=0}^k {}_k C_j\, (1-r)^{n j} e^{2\pi i \epsilon \alpha j} r^{n(k-j)}\,,
\eeq
so performing the Fourier transform we get
\beq
S^{1^\epsilon 1^\epsilon \ldots 1^\epsilon}_n(r;q)=\frac{1}{1-n}\log\frac{\sum_{j=0}^k {}_k C_j\, (1-r)^{n j} \delta_{q,\epsilon j} r^{n(k-j)}}{\left[ \sum_{j=0}^k {}_k C_j\, (1-r)^{j} \delta_{q,\epsilon j} r^{(k-j)}\right]^n}\,,
\label{skk}
\eeq
thus for a particular value of the charge we have 
\beq
S^{1^\epsilon 1^\epsilon \ldots 1^\epsilon}_n(r;\epsilon j)
=\frac{1}{1-n}\log\frac{ {}_k C_j\, (1-r)^{n j}  r^{n(k-j)}}{\left[{}_k C_j\, (1-r)^{j} r^{(k-j)}\right]^n}=\log {}_k C_j\,.
\label{skkc}
\eeq
\begin{figure}[t]
	\begin{subfigure}{.5\textwidth}
	\includegraphics[width=\linewidth]{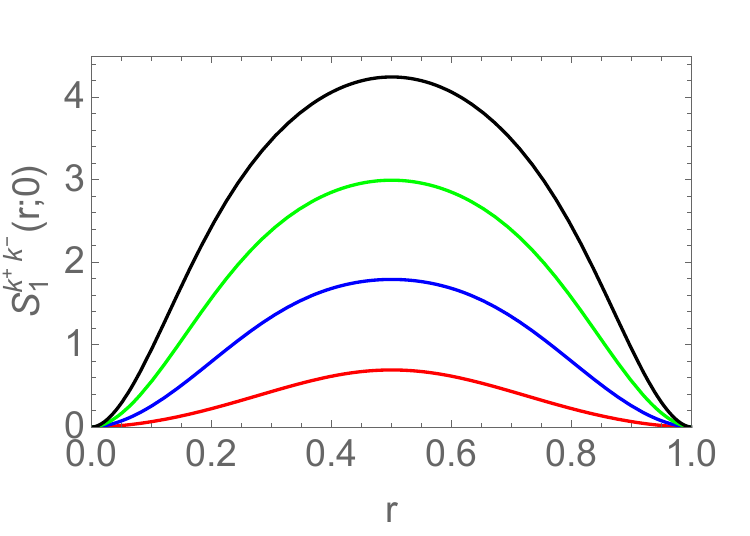}
	\end{subfigure}
	\begin{subfigure}{.5\textwidth}
	\includegraphics[width=\linewidth]{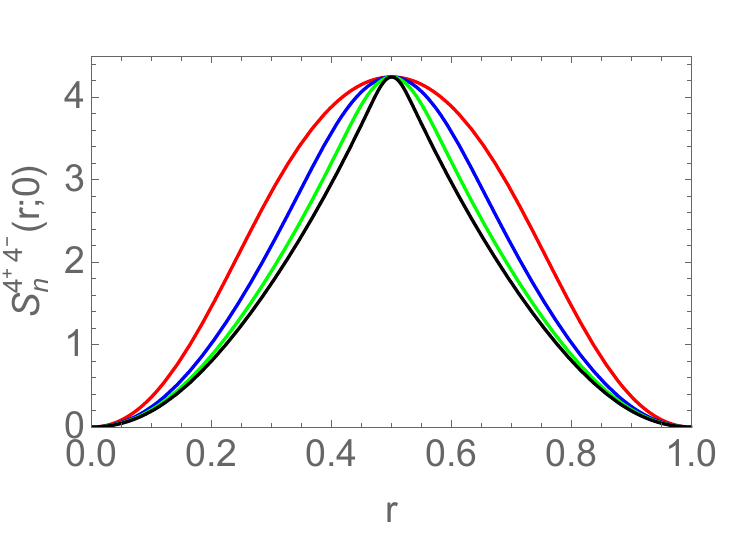}
    \end{subfigure}
    \caption{{ Symmetry resolved entropies of various qubit states in the zero charge sector as functions of $r$.} Left: The symmetry resolved von Neumann entropy of the charge zero sector for states of equal numbers $k$ of identical positively and negatively charged particles. In the figure $k=1,2,3,4$ giving larger entropy for higher $k$. The maxima at $r=1/2$ are $\log 2, \log 6, \log 20$ and $\log 70$, that is $\log(2k)!-2\log(k!)$ which counts the number of distinct arrangements of two groups of $k$ identical particles. Right: The symmetry resolved R\'enyi entropy of the charge zero sector of a state consisting of four identical positively and four identical negatively charged excitations for $n=2,4,8,20$. The larger $n$ is, the more sharply peaked at $r=1/2$ the functions become. The value at $r=1/2$ is $\log 70$, independent of $n$.}
    \label{fig1}
\end{figure}
In this case the SREE tells us about the number of equally likely configurations which produce a charge $\epsilon j$ in region $A$ and is independent of $n$. 
Many other configurations can be considered, all of which produce different results, with similar interpretations. For instance, for a state with one positively and one negatively charged particle, the Fourier transform of the function
\beqa
&& (r^n+e^{2\pi i \alpha}(1-r)^n) (r^n+e^{-2\pi i \alpha}(1-r)^n)\,,
\eeqa 
gives the simple formula
\beq
S^{1^+ 1^-}_n(r;q)=\frac{1}{1-n}\log\frac{(r^{2n} +(1-r)^{2n})\delta_{q,0}+r^n (1-r)^n (\delta_{q,1}+\delta_{q,-1})}{\left[ (r^{2} +(1-r)^{2})\delta_{q,0}+r (1-r) (\delta_{q,1}+\delta_{q,-1})\right]^n}\,,
\label{skkk}
\eeq
and
\begin{eqnarray}
S^{1^+ 1^-}_n(r;0)=\frac{1}{1-n} \log \frac{r^{2n}+(1-r)^{2n}}{(r^2+(1-r)^2)^n}\,,\quad S^{1^+ 1^-}_n(r;\pm 1)=0\,.
\label{noequ}
\end{eqnarray}
In this case the $q=0$ result is $n$-dependent and gives a non-trivial symmetry resolved von Neumann entropy:
{
\begin{eqnarray}
S^{1^+ 1^-}_1(r;0)=\log (r^2+(1-r)^2)-\frac{r^2 \log r^2+(1-r)^2 \log (1-r)^2}{r^2+(1-r)^2}\,,\quad S^{1^+ 1^-}_1(r;\pm 1)=0\,.
\label{noequ2}
\end{eqnarray}}
In this example the SREE of the $q=0$ sector is non-trivial as there are now two possible configurations that we can associate with such a charge, namely both particles being in region $A$ and no particle being in region $A$. Thus there is a difference in the SREEs of states involving two particles with the same or distinct charges, even for the  simple states considered here. Additional examples are presented in Fig.~\ref{fig1}.

It is worth noting that all formulae in this section are in agreement with those in Section 4 if we identify the function (\ref{funk}) with the ground-state partition function $\mathcal{Z}_n^0(q)$. Therefore, the study of qubit states provides a neat application of the general results of Section 4 for the case of a trivial, unentangled, ground state.

{ Because of the simplicity and explicit nature of all the formulae in this subsection, it is now possible to compute precisely the two contributions to the total  von Neumann entropy that are often discussed in the literature, namely the configurational entropy and the number entropy \cite{GS,german3}. Calling $S_1^{\Psi}(r)$ the total von Neumann entropy of the state $|\Psi\ket$ we can write
\beq 
S_1^{\Psi}(r)=\sum_{q} (p(q)S_1^\Psi(q;r)-p(q)\log p(q))\,,
\label{68}
\eeq 
where $p(q):=\mathcal{Z}_1^\Psi(r;q)$, that is the symmetry resolved partition function of the state for $n=1$ and the term $\sum_q p(q)\log p(q)$ is the number entropy. This represents the probability of obtaining the value $q$ when measuring the charge. 

It is easy to work out an explicit example and see the features of these two contributions. For the same state of two excitations of distinct charges (\ref{noequ})-(\ref{noequ2}) we have that 
\beq 
p(0)=r^2+(1-r)^2,\qquad p(\pm 1)=r(1-r)\,,
\eeq
so that the number entropy is simply
\beq 
(r^2+(1-r)^2)\log (r^2+(1-r)^2)+2r(1-r)\log(r(1-r))\,,
\eeq 
and the configuration entropy is
\beq 
(r^2+(1-r)^2)S^{1^+ 1^-}_1(r;0)\,,
\eeq 
with $S^{1^+ 1^-}_1(r;0)$ given by (\ref{noequ2}). It is very easy to evaluate (\ref{68}) with these contributions and to recover the known formula for the total von Neumann entropy of a state of two distinct excitations $-2r \log r-2(1-r)\log(1-r)$ as found in \cite{excited,excited2}. We note also that the number entropy takes its maximum (absolute) value $3/2\log 2$ at $r=1/2$, and that it can itself be considered a measure of entanglement, as discussed for other examples in \cite{ref1,ref2}.
}
\medskip

{  We close this subsection by noting that the entropy formulae for qubit states considered here do not have the property of equipartition, that is, they depend explicitly on the charge sector as we see for instance from Eq. (\ref{noequ}). This is no contradiction as the property of equipartition \cite{german3} is typically a leading order property (for instance in \cite{german3} it holds for small magnetization). In the case of qubit states we have exact formulae rather than leading order expressions, thus they depend on the charge. Indeed, they provide probably the simplest example where such a dependence can be easily shown.}

\section{Conclusions}
\label{theend}
In this paper we have computed the symmetry resolved entanglement entropy and its moments for zero-density excited states. These are defined as excited states consisting of a finite number of excitations above the ground state in a scaling limit where both the volume of the system and the volume of each subsystem are taken to infinity,  keeping their ratio constant. 
\medskip

It is known from previous work \cite{excited,excited2,excited3,excited4,ali1,ali2,ali3,ali4,ali5,ali6} that the difference between the entanglement entropy of the excited state and that of the ground state, also known as excess entropy, takes an extremely simple and universal form for non-interacting 1+1D QFTs and also for certain highly excited states of CFT \cite{Capizzi_2020}. 
Since this excess entanglement represents the extra contribution to entanglement of an excited state above a non-trivially entangled ground state it has also been realised that the same extra contribution is obtained when the ground state is trivial. For this reason both a free QFT and a qubit picture lead to the same results, even if the underlying theories are extremely different.  Finally, it has also been shown that the results extend to free bosons in any dimension \cite{excited3} and more generally, to any situations where excitations are localised, in the sense that either the correlation length or the De Broglie wave length of the excitations are small compared to region sizes. 

The results in this paper are extensions of the work we have just summarised and hold under the same assumptions. However, while the work above dealt with the excess entanglement, the present work deals with the SREE and its associated charged moments (that is, the Fourier transform of the associated partition function). It turns out that the ratio of charged moments between the excited and ground states, takes a universal form which is a simple generalisation of the results for the excess entropy. While this generalisation is very natural and not difficult to obtain from previous work, what is perhaps more novel and surprising is that from the ratio of charged moments, provided these are well-defined in the infinite volume ground state, it is also possible to obtain exact expressions for the SREE of the excited states. These expressions can be written solely in terms of the SREE and symmetry resolved partition function of the ground state, thus are again very widely applicable. In particular, when qubit states are considered, totally explicit formulae for the SREEs can be obtained in this manner. Although the focus of this paper has been on models with $U(1)$ symmetry, we expect analogous formulae to hold for other symmetries, both continuous and discrete. 

\medskip

There are various problems that we plan to address in the near future: extending our results to interacting and higher-dimensional theories as well as providing numerical verification of our formulae.  We will present these results in \cite{partII}. Looking further, we would like to extend these results to the symmetry resolved negativity and to study finite-volume corrections employing the form factor techniques presented here. It would also be interesting to investigate the correlation functions of CTFs in infinite volume for free theories, along the lines of \cite{Isingb, FB}.

\medskip

\noindent {\bf Acknowledgements:} We would like to thank Benjamin Doyon for useful discussions. Luca Capizzi thanks ERC for support under Consolidator grant number 771536 (NEMO). Cecilia De Fazio thanks the Engineering and Physical Sciences Research Council for financial support under EPSRC Grant EP/V031201/1. Michele Mazzoni is grateful for funding under the EPSRC Mathematical Sciences Doctoral Training Partnership EP/W524104/1. Luc\'ia Santamar\'ia-Sanz is grateful to the Spanish Government for funding under the FPU-fellowships program FPU18/00957, the FPU Mobility subprogram EST19/00616, and MCIN grant PID2020-113406GB-I0.

\appendix 
\section{Complex Free Boson Computation}
In this Appendix we present the form factor computation of the ratio of charged moments in detail, focusing on the complex free boson theory. 
\subsection{Single-Particle Excited States}
Once the two-point function in (\ref{expansion}) is obtained we can compute it by inserting a sum over a complete set of states between the $U(1)$ fields as follows:
\begin{eqnarray}
\label{matrixelem}
&&{} _L^n\bra {1}^+ |\TT^\alpha_n (0) \Tilde{\TT}^\alpha_n (\ell) |{1}^+\ket_L^n  = \sum_{\{ N^+ \}} \,\sum_{ \{ M^+\}} A_n^* (\{ N^+ \}) \; A_n (\{ M^+ \}) \prod_{p=1}^{n} \sum_{m^\pm=0}^\infty \sum_{\{ J^\pm \}} \prod_{j=1}^{m^+}\; \prod_{r=1}^{m^-}\; \frac{1}{m^+!\, m^-!}\nonumber \\
&&  \times  {}^n_{p,L}\bra {0}| [ \textfrak{a}_{p} \; (\theta) ]^{N^+_p} \;  \TT_{p+\alpha} (0)  \textfrak{a}_{p}^\dag (\theta^+_j)\textfrak{b}_{p}^\dag (\theta^-_r) |{0}\ket_{p,L}^n \; \times \; {}^n_{p,L}\bra {0}|\textfrak{a}_{p} (\theta^+_j)\textfrak{b}_{p} (\theta^-_r) \TT_{-p-\alpha} (\ell)  \; [ \textfrak{a}_{p}^\dag (\theta) ]^{M^+_p} \;  |{0}\ket_{p,L}^n \;,\nonumber\\
\end{eqnarray} 
and similarly for the $|{1}^-\ket_L^n $ case. Since the matrix elements involved are related to the infinite-volume form factors, we can rewrite the previous expression up to exponentially decaying corrections as
\begin{eqnarray}
&&{} _L^n\bra {1}^+ |\TT^\alpha_n (0) \Tilde{\TT}^\alpha_n (\ell) |{1}^+\ket_L^n  = \sum_{\{ N^+ \}} \,\sum_{ \{ M^+\}} A_n^* (\{ N^+ \}) \; A_n (\{ M^+ \}) \prod_{p=1}^{n} \sum_{m^\pm=0}^\infty \sum_{\{ J^\pm \}} \; \frac{1}{m^+!\, m^-!} \\
&& \times \frac{e^{i \ell \left(\sum_{j=1}^{m^+} P(\theta^+_j)+\sum_{r=1}^{m^-} P(\theta^-_r)-M_p^+ P(\theta)\right)}}{\sqrt{L E(\theta)}^{N_p^++M_p^+} \prod_{j=1}^{m^+} L E(\theta^+_j)\, \prod_{r=1}^{m^-} L E(\theta^-_r)} F_{N_p^++m^++m^-}^{n,p}(\theta^+_1 \dots \theta^+_{m^+},\hat{\theta}, \dots \hat{\theta}, \theta^-_1 \dots \theta^-_{m^-}) \nonumber\\
&&\times F_{M_p^++m^++m^-}^{n-p,n}(\theta \dots \theta,\hat{\theta}^-_1 \dots \hat{\theta}^-_{m^-}, \hat{\theta}^+_1 \dots \hat{\theta}^+_{m^+})\,,\nonumber 
\end{eqnarray} 
being $\hat{\theta}^\pm_j=  \theta^\pm_j+ i \pi$, $E(\theta)= m \cosh \theta$ and $P(\theta^\pm_j), P(\theta)$ given by the Bethe-Yang quantisation condition (\ref{quan}). The complete formula for the form factors above was given in \cite{excited2} and they can be fully expressed as sums of products of two-particle form factors. They are non-vanishing for   $N_p^+=M_p^+=m^+-m^-$ and zero otherwise. 

If the same intermediate rapidity $\theta^+_j$ is paired up in the Wick-contraction sense with the rapidity of the excited state state $\theta$ from the in- an out-states, the dominant contribution in the form factor product will come from kinematic poles. In other words, if $\theta^+_j \sim \theta$ two-particle form factors will appear as follows:
\begin{eqnarray*}
&&\hspace{-15pt}F_{N_p^++m^++m^-}^{n,p}(\theta^+_1 \dots \theta^+_{m^+},\hat{\theta} \dots \hat{\theta},\dots) \sim N_p^+ f_{p+\alpha}^n(\theta^+_j-\hat{\theta})\nonumber\\
&&\times F_{N_p^++m^++m^--2}^{n,p}(\theta^+_1 \dots \theta^+_{j-1}\theta^+_{j+1} \dots \theta^+_{m^+},\hat{\theta} \dots \hat{\theta}\dots)\\
&&\hspace{-15pt}F_{M_p^++m^++m^-}^{n-p,n}(\theta \dots \theta, \dots \hat{\theta}^+_1 \dots \hat{\theta}^+_{m^+}) \sim M_p^+ f_{n-(p+\alpha)}^n(\hat{\theta}-\theta^+_j)\nonumber\\
&&\times F_{M_p^++m^++m^--2}^{n-p,n}(\theta \dots \theta, \dots \hat{\theta}^+_1 \dots \hat{\theta}^+_{j-1}\hat{\theta}^+_{j+1} \dots \hat{\theta}^+_{m^+}),
\end{eqnarray*}
where the number of $\hat{\theta}$ ($\theta$) in the arguments of the form factors in the right-hand side term are now $N_p^+-1$ ($M_p^+-1$). The main property of the matrix elements in (\ref{matrixelem}) that determines the final formula for (\ref{expansion}) is the infinite volume limit of the terms such as
\beqa
&& \sum_{J^+ \in \mathbb{Z}} \frac{f_{p+\alpha}^n(\theta^+ - \hat{\theta})f_{n-p-\alpha}^n(\theta-\hat{\theta}^+) e^{i\ell(P(\theta^+)-P(\theta))}}{\cosh\theta \cosh\theta^+} \sim \nonumber\\
&& (mL)^2 \sum_{J_i^+ \in \mathbb{Z}} \frac{\sin^2\frac{\pi(p+\alpha)}{n}}{\pi^2} \frac{e^{2\pi i r (J^+-I+\frac{p+\alpha}{n})}}{ (J^+-I+\frac{p+\alpha}{n})^2}=(mL)^2 g_{p+\alpha}^n(r)\,,
\eeqa
with $g_{p+\alpha}^n(r)$ the functions defined in (\ref{gfun}) and 
the indices $J^+, I$ are integers resulting from the quantisation conditions of the rapidities of intermediate states (\ref{quan}) and of the rapidity of the physical one-particle state $P(\theta)=2\pi I$ with $I\in \mathbb{Z}$. We can proceed in an analogous way for (\ref{expansion1}) obtaining $(mL)^2 g_{-(p+\alpha)}^n(r)$ as the leading contribution. 

Once all possible contractions with a rapidity of the excited in- and out- state have been carried out, the leading large-volume contribution from the summation over the quantum number $J^+$ is of order $L^0$ and comes from terms with $N=M$, as shown in Appendix B of \cite{excited2}. It can be written as
\begin{eqnarray*}
&&\!\!\!{} _L^n\bra {1}^+ |\TT^\alpha_n (0) \Tilde{\TT}^\alpha_n (\ell) |{1}^+\ket_L^n  = \sum_{\{ N^+ \}} \, |A_n (\{ N^+ \})|^2 \prod_{p=1}^{n} N_p^+! \left[g^n_{p+\alpha}(r)\right]^{N_p^+} \prod_{q^+=0}^{\infty}\prod_{m^-=0}^{\infty} \frac{1}{q^+!\, m^-!} \sum_{\{ J^\pm \} \in \mathbb{Z}}  \\
&&\!\!\! \times \frac{e^{i \ell \left(\sum_{j=1}^{q^+} P(\theta^+_j)+\sum_{r=1}^{m^-} P(\theta^-_r) \right)}}{\prod_{j=1}^{q^+} L^2 E(\theta^+_j)\prod_{r=1}^{m^-} L^2 E(\theta^-_r)} F_{q^++m^-}^{p,n} (\theta^+_1 \dots \theta^+_{q^+},\theta^-_1 \dots \theta^-_{m^-})F_{q^++m^-}^{n-p,n} (\hat{\theta}^+_1\dots\hat{\theta}^+_{q^+},\hat{\theta}^-_1\dots\hat{\theta}^-_{m^-})
\end{eqnarray*}
with $q^+=m^+-N_p^+$. Dividing by the finite-volume vacuum two-point function in the given sector ${} _L^p\bra 0 |\TT_p (0) \Tilde{\TT}_p (\ell) |0 \ket_L^p$ we obtain the formula (\ref{1par}) for the ratio of moments of the SREE for a one excitation state. 

\subsection{Free Boson ($k=1$,$n=2$)}
In this section we work out an example in detail. Consider a single particle excited state consisting of a complex boson excitation above the ground state. The relevant state is
\beq
(a_1^+)^\dagger(\theta)(a_2^+)^\dagger(\theta)|0\ket_L^2=\frac{1}{2}(-\textfrak{a}_1^\dagger(\theta)+\textfrak{a}_2^\dagger(\theta))(\textfrak{a}_1^\dagger(\theta)+\textfrak{a}_2^\dagger(\theta))|0\ket_L^2\,,
\eeq
so, comparing to the state (\ref{oldnotationstate})
we have that $A_2(2,0)=-A_2(0,2)=-\frac{1}{2}$. Thus
\beqa
M_n^{1^+}(r;\alpha)&=&\frac{2!}{4} (g_{1+\alpha}^2(r)^2+g_{2+\alpha}^2(r)^2)\nonumber\\
&=&
\frac{1}{2}\left(1-r+r e^{\pi i(1+\alpha)}\right)^2+ \frac{1}{2}\left(1-r+r e^{\pi i(2+\alpha)}\right)^2\nonumber\\
&=& \frac{1}{2}((1-r)^2+ r^2 e^{2\pi i \alpha}-2 r(1-r)e^{i\pi\alpha}+(1-r)^2+ r^2 e^{2\pi i \alpha}+2r(1-r)e^{i\pi\alpha}) \nonumber\\
&=&(1-r)^2+e^{2\pi i \alpha} r^2\,.
\eeqa 

\subsection{Multi-Particle Excited States}
Below, we describe in detail the computation of the ratio of moments of the SREE for a state consisting of $k$ particle excitations with equal rapidities and charge signs. These states have the form:
\begin{equation}
|k^\pm \ket_L^n= \prod_{j=1}^n (a_j^\pm)^\dagger (\theta)|0\ket_L^n = \frac{1}{(\sqrt{k!})^n} \, \sum_{\{N^\pm\}} D_n({\{N^\pm\}}) \prod_{p=1}^n \left[ \left(\tilde{a}_p^\pm\right)^\dagger(\theta)\right]^{N_p^\pm} |0\ket_{L}^n
\end{equation}
where in the last equality we have used the expression of the creation operators in the diagonal basis described in (\ref{rels}). The two point function would be:
\begin{eqnarray}
&& \!\!\!\!\!  \!\!\!\!\!  {} _L^n\bra {k}^\pm |\TT^\alpha_n (0) \Tilde{\TT}^\alpha_n (\ell) |{k}^\pm \ket_L^n  = \frac{1}{(k!)^n}\sum_{\{ N^\pm \}} \,\sum_{ \{ M^\pm\}} D_n(\{ N^+ \}) \; D_n^* (\{ M^+ \}) \prod_{p=1}^{n} \sum_{m^\pm=0}^\infty \sum_{\{ J^\pm \}} \prod_{j=1}^{m^+}\; \prod_{r=1}^{m^-}\; \frac{1}{m^+!\, m^-!}\nonumber \\
&& \!\!\!\!\!  \!\!\!\!\!  \times  {}^n_{p,L}\bra {0}| [ \tilde{a}^\pm_{p} \; (\theta) ]^{N^\pm_p} \;  \TT_{p+\alpha} (0)  \textfrak{a}_{p}^\dag (\theta^+_j)\textfrak{b}_{p}^\dag (\theta^-_r) |{0}\ket_{p,L}^n  \; {}^n_{p,L}\bra {0}|\textfrak{a}_{p} (\theta^+_j)\textfrak{b}_{p} (\theta^-_r) \TT_{-p-\alpha} (\ell)  \; [ (\tilde{a}^\pm_{p})^\dag (\theta) ]^{M^\pm_p} \;  |{0}\ket_{p,L}^n \;,\nonumber\\
\end{eqnarray}
where we have inserted a complete set of states between the two twist fields. Employing the relation between these matrix elements and the finite volume form factors and the action of the translation operator on energy states, we get:
\begin{eqnarray}
&&{} _L^n\bra {k}^\pm |\TT^\alpha_n (0) \Tilde{\TT}^\alpha_n (\ell) |{k}^\pm\ket_L^n  = \frac{1}{(k!)^n} \sum_{\{ N^+ \}} \,\sum_{ \{ M^+\}} D_n (\{ N^+ \}) \; D_n^* (\{ M^+ \}) \prod_{p=1}^{n} \sum_{m^\pm=0}^\infty \sum_{\{ J^\pm \}} \; \frac{1}{m^+!\, m^-!} \nonumber\\
&& \times \frac{e^{i \ell \left(\sum_{j=1}^{m^+} P(\theta^+_j)+\sum_{r=1}^{m^-} P(\theta^-_r)-M_p^\pm P(\theta)\right)}}{\sqrt{L E(\theta)}^{N_p^\pm+M_p^\pm} \prod_{j=1}^{m^+} L E(\theta^+_j)\, \prod_{r=1}^{m^-} L E(\theta^-_r)} F_{N_p^\pm+m^++m^-}^{n,p}(\theta^+_1 \dots \theta^+_{m^+},\hat{\theta}, \dots \hat{\theta}, \theta^-_1 \dots \theta^-_{m^-}) \nonumber\\
&&\times F_{M_p^++m^++m^-}^{n-p,n}(\theta \dots \theta,\hat{\theta}^-_1 \dots \hat{\theta}^-_{m^-}, \hat{\theta}^+_1 \dots \hat{\theta}^+_{m^+})\,.\nonumber 
\end{eqnarray}
Once all possible intermediate rapidities have been paired up with the same rapidity of the excited state in both form factors and the contribution of the ground state factored out as explained in previous sections, the leading large volume contribution of the ratio of moments can be written as:
\begin{eqnarray}
M^{k^\pm}_n(r;\alpha)&=& \frac{1}{(k!)^n} \sum_{\{N^\pm\}}|D_n(\{N^\pm\})|^2 \prod_{p=1}^n N_{p}^\pm!\,\, \left( g^n_{\pm(p+\alpha)} (r) \right)^{ N_{p}^\pm} \nonumber\\
&=& \sum_{p=0}^k \left[\left(\begin{array}{c}
k\\
p
\end{array}\right) r^p (1-r)^{k-p}\right]^n e^{\pm 2\pi i \alpha p}\,.
\end{eqnarray}
Nevertheless, for $k$-particle excitations with distinct rapidities there could be two different cases:
\begin{itemize}
    \item $k$-particle excitations with distinct rapidities but equal charge sign
    \item $k$-particle excitations with distinct rapidities and charge sign
\end{itemize}
We can summarize the computations for both cases if we consider the following ansatz for the excited state:
\begin{equation}
|1^{\epsilon_1} \, 1^{\epsilon_2} \dots 1^{\epsilon_k}\ket_L^n = \prod_{i=1}^k \prod_{j=1}^n (a_j^{\epsilon_i})^{\dagger} (\theta_i) |0\ket_L^n = \sum_{\{ N^\pm \}} C_n(\{ N^\pm \}) \prod_{s=1}^k \prod_{p=1}^n [\textfrak{a}_p^\dagger(\theta_s)]^{N_{p,s}^+}[\textfrak{b}_p^\dagger(\theta_s)]^{N_{p,s}^-} |0\ket_L^n
\end{equation}
where each $\epsilon_i$ with $i=1,\dots, k$ could be $+$ or $-$. If we consider $k$-particle excitations with distinct rapidities but equal charge sign $+$ ($-$) then all the $\epsilon_i$ are the same sign and $N_{p,s}^-$ ($N_{p,s}^+$) vanish. 
Pairing up the intermediate rapidities with the same rapidity of the excited state in both form factors means that 
\begin{eqnarray*}
N_{p,s}^- + m^+ &=& N_{p,s}^++m^-\\
m^-+M_{p,s}^+ &=& m^+ +M_{p,s}^-
\end{eqnarray*}
in order for the matrix elements arising in the intermediate steps to be non-vanishing. Subtracting the contribution of the ground state, the leading large volume contribution to the ratio of moments can be written as:
\begin{eqnarray}
M^{1^{\epsilon_1} \, 1^{\epsilon_2} \dots 1^{\epsilon_k}}_n(r;\alpha)= \prod_{s=1}^k \left[\sum_{\{N^\pm\}}|C_n(\{N^\pm\})|^2 \prod_{p=1}^n N_{p,s}^+!\,\,N_{p,s}^-!\, \,[ g^n_{p+\alpha} (r)]^{ N_{p,s}^+} \, [g^n_{-p-\alpha} (r)]^{ N_{p,s}^-}\right]. 
\end{eqnarray}
Notice that if we study $k$-particle excitations with distinct rapidities but equal charge sign this last expression reduces to the following one:
\begin{eqnarray}
M^{1^{\pm} \, 1^{\pm} \dots 1^{\pm}}_n(r;\alpha)= \prod_{s=1}^k \left[\sum_{\{N^\pm\}}|C_n(\{N^\pm\})|^2 \prod_{p=1}^n N_{p,s}^\pm!\, \,[ g^n_{\pm(p+\alpha)} (r)]^{ N_{p,s}^\pm}\right]\,,
\end{eqnarray}
with the conditions $m^{\mp}+M_{p,s}^\pm  = m^\pm = N_{p,s}^\pm +m^\mp$ for the $\pm$ sign state.

\subsection{Free Boson ($k=2$,$n=2$)}
On the one hand, consider the following two-particle excited states with distinct rapidities:
\begin{eqnarray}
|1^+ 1^+\ket_L^2 &=& (a_1^+)^\dagger(\theta_1)(a_1^+)^\dagger(\theta_2)(a_2^+)^\dagger(\theta_1)(a_2^+)^\dagger(\theta_2)|0\ket_L^2=\frac{1}{4}\left([\textfrak{a}_1^\dagger(\theta_1)]^2 \, [\textfrak{a}_1^\dagger(\theta_2)]^2\right. \nonumber\\
&+&\left. [\textfrak{a}_2^\dagger(\theta_1)]^2 \, [\textfrak{a}_2^\dagger(\theta_2)]^2-[\textfrak{a}_1^\dagger(\theta_2)]^2 \, [\textfrak{a}_2^\dagger(\theta_1)]^2-[\textfrak{a}_1^\dagger(\theta_1)]^2 \, [\textfrak{a}_2^\dagger(\theta_2)]^2\right)\, |0\ket_L^2\,,\\
|1^+ 1^-\ket_L^2 &=& (a_1^+)^\dagger(\theta_1)(a_1^-)^\dagger(\theta_2)(a_2^+)^\dagger(\theta_1)(a_2^-)^\dagger(\theta_2)|0\ket_L^2=\frac{1}{4}\left([\textfrak{a}_1^\dagger(\theta_1)]^2 \, [\textfrak{b}_1^\dagger(\theta_2)]^2\right. \nonumber\\
&+&\left. [\textfrak{a}_2^\dagger(\theta_1)]^2 \, [\textfrak{b}_2^\dagger(\theta_2)]^2-[\textfrak{a}_2^\dagger(\theta_1)]^2 \, [\textfrak{b}_1^\dagger(\theta_2)]^2-[\textfrak{a}_1^\dagger(\theta_1)]^2 \, [\textfrak{b}_2^\dagger(\theta_2)]^2\right)\, |0\ket_L^2\,.
\end{eqnarray}
Hence the ratio of the moments of the SREE for these excited states are given by
\begin{eqnarray}
M_2^{1^+ 1^+}(r;\alpha)&=&\frac{1}{4} \left(  [g_{1+\alpha}^2(r)]^4+  [g_{2+\alpha}^2(r)]^4+ 2 \, [g_{1+\alpha}^2(r)]^2\,  [g_{2+\alpha}^2(r)]^2)\right)\nonumber\\ 
&=&\left( (1-r)^2+r^2 e^{2 \pi i \alpha}\right)^2 \,,\\
M_2^{1^+ 1^-}(r;\alpha)&=&\frac{1}{4} \left(  [g_{1+\alpha}^2(r)]^2\, [g_{-1-\alpha}^2(r)]^2 + [g_{2+\alpha}^2(r)]^2\, [g_{-2-\alpha}^2(r)]^2+  [g_{1+\alpha}^2(r)]^2\,  [g_{-2-\alpha}^2(r)]^2 \right.\nonumber\\ 
&+& \left [g_{-1-\alpha}^2(r)]^2\, [g_{2+\alpha}^2(r)]^2\, \right)=\left( (1-r)^2+r^2 e^{2 \pi i \alpha}\right)\left( (1-r)^2+r^2 e^{-2 \pi i \alpha}\right) \,.
\end{eqnarray}
On the other hand, consider the following two-particle excited state with coinciding rapidity:
\beqa
 |2^-\ket_L^2 &=& [(a_1^-)^\dagger(\theta)]^2\,  [(a_2^-)^\dagger(\theta) ]^2 |0\ket_L^2 \nonumber \\ &=&\frac{1}{8} \left( [\textfrak{b}_1^\dagger(\theta)]^4 +[\textfrak{b}_2^\dagger(\theta)]^4 -2 [\textfrak{b}_1^\dagger(\theta)]^2\, [\textfrak{b}_2^\dagger(\theta)]^2 \right) |0\ket_L^2\,.
\eeqa
Hence the ratio of moments of the SREE for this excited states can be written as
\beqa
M_2^{2^-}(r;\alpha)&=&\frac{1}{8^2} ( 4!\, [g_{-1-\alpha}^2(r)]^4+ 4!\, [g_{-2-\alpha}^2(r)]^4+ 4^2 \, [g_{-1-\alpha}^2(r)]^2\,  [g_{-2-\alpha}^2(r)]^2)\nonumber\\ 
&=&(1-r)^4+4\, r^2 (1-r)^2 e^{-2\pi i \alpha}+ e^{-4\pi i \alpha} r^4\,.
\eeqa 

\section{Complex Free Fermion Computation}
In this Appendix we present the form factor computation of the ratio of charged moments in detail, focusing on the complex free fermion theory. 
\subsection{Single-Particle Excited States} 
Below, we present the explicit computation of the fermionic two-point function in an excited state consisting of a single positively-charged particle. Thanks to the factorisation \eqref{TTfactorizesector_fermion}, the latter can be cast as:
\begin{align}
&_L^n\bra {1}^+ |\TT^\alpha_n (0) \Tilde{\TT}^\alpha_n (\ell) |{1}^+\ket_L^n \nonumber \\ = &\prod_{p=-\frac{n-1}{2}}^{\frac{n-1}{2}}{}^n_{p,L}\bra {0}| \textfrak{a}_{p} \; (\theta)  \;  \TT_{p+\alpha} (0) \; \TT_{-p-\alpha} (\ell)  \; \textfrak{a}_{p}^\dag (\theta)  \;  |{0}\ket_{p,L}^n \nonumber\\
= &\prod_{p=-\frac{n-1}{2}}^{\frac{n-1}{2}}\sum_{s=0}^{\infty}\sum_{\{J_i^\pm\}}\frac{1}{s!(s+1)!}{}^n_{p,L}\bra {0}| \textfrak{a}_{p} \; (\theta)  \;  \TT_{p+\alpha} (0) \; \textfrak{a}_{p}^\dag (\theta_1)\dots \textfrak{a}_{p}^\dag (\theta_{s+1})\textfrak{b}_{p}^\dag (\theta_{s+2})\dots\textfrak{b}_{p}^\dag (\theta_{2s+1})|{0}\ket_{p,L}^n \nonumber\\ &\times {}^n_{p,L}\bra {0}| \textfrak{a}_{p} \; (\theta_1)  \;\dots \textfrak{a}_{p} (\theta_{s+1})\textfrak{b}_{p} (\theta_{s+2})\dots\textfrak{b}_{p} (\theta_{2s+1}) \TT_{-p-\alpha} (0) \; \textfrak{a}_{p}^\dag (\theta) |{0}\ket_{p,L}^n e^{i\ell (\sum_{i=1}^{2s+1}P(\theta_i)-P(\theta))} \nonumber \\
& =\prod_{p=-\frac{n-1}{2}}^{\frac{n-1}{2}}\sum_{s=0}^{\infty}\sum_{\{J_i^\pm\}} \frac{\abs{{F^{p+\alpha,n}_{2s+2}(\theta_1,\dots,\theta_{s+1};\theta + i\pi,\theta_{s+2},\dots,\theta_{2s+1})}}^2}{s!(s+1)!LE(\theta)\prod_{i=1}^{2s+1}(\theta_i)LE(\theta_i)}e^{i\ell (\sum_{i=1}^{2s+1}P(\theta_i)-P(\theta))}\,,
\end{align}
where the resolution of the identity is inserted in such a way as to preserve the total charge of the one-particle state and the Bethe quantum numbers $\{J_i^\pm\}$ are defined as in \eqref{quan}. Notice that since the excitations are fermionic, one could either have $J_i^\pm \in \mathbb{Z}$ or $J_i^\pm \in \mathbb{Z} + \frac{1}{2}$: for the sake of simplicity we will consider the case where these numbers are integer. 

The non-vanishing contributions in the $L \to +\infty$ limit come from the terms in the previous expression in which the rapidity of the excited state is contracted with $\theta_i$, $i=1,\dots, s+1$ in both form factors. The $s+1$ possible contractions in $F^{p+\alpha,n}_{2s+2}$ give rise to:
\begin{align}
&F^{p+\alpha,n}_{2s+2}(\theta_1,\dots,\theta_{s+1};\hat{\theta},\theta_{s+2},\dots,\theta_{2s+1})  \nonumber \\
& \sim f^n_{p+\alpha}(\theta_i - \hat{\theta})\,F^{p+\alpha,n}_{2s}(\theta_1,\dots,\Check{\theta_i},\dots,\theta_{s+1};\theta_{s+2},\dots,\theta_{2s+1})
\end{align}
where around the pole:
\beq
\label{fermionic_two-particle_ff_simple_pole}
f^n_{p+\alpha}(\theta_i - \hat{\theta}) \underset{\theta \sim \theta_i}{\sim} \frac{mL \sin{\frac{\pi(p+\alpha)}{n}}\cosh{\theta}\,e^{\frac{i\pi(p+\alpha)}{n}}}{\pi(J_i^+ - I + \frac{p+\alpha}{n})}
\eeq
so that, considering also the contraction coming from the conjugate form factor, we can separately perform the $s+1$ summations over the quantum numbers $J_i^+$ as
\begin{align}
&\sum_{J_i^+ \in \mathbb{Z}} \frac{\abs{f^n_{p+\alpha}(\theta_i - \hat{\theta})}^2 e^{i\ell(P(\theta_i)-P(\theta))}}{L m \cosh{\theta}\,L m \cosh{\theta_i}} \nonumber \\
\underset{\theta \sim \theta_i}{\sim} &\sum_{J_i^+ \in \mathbb{Z}} \frac{\sin^2\frac{\pi(p+\alpha)}{n}e^{2 \pi i r \left(J_i^+ - I + \frac{p+\alpha}{n}\right)}}{\pi^2(J_i^+ - I + \frac{p+\alpha}{n})^2} = g^n_{p+\alpha}(r)\,.
\end{align}
We therefore obtain, in the limit $L , \ell \to \infty$ and for fixed $r$:
\begin{align}
&_L^n\bra {1}^+ |\TT^\alpha_n (0) \Tilde{\TT}^\alpha_n (\ell) |{1}^+\ket_L^n \nonumber \\ =
& \prod_{p=-\frac{n-1}{2}}^{\frac{n-1}{2}}g^n_{p+\alpha}(r)\sum_{s=0}^{\infty}\frac{1}{(s!)^2}\sum_{\{J_i^\pm\}} \abs{{F^{p+\alpha,n}_{2s}(\theta_1,\dots,\theta_s;\beta_1,\dots,\beta_s)}}^2\frac{e^{i\ell \sum_{i=1}^s(P(\theta_i)+P(\beta_i))}}{\prod_{i=1}^{s}LE(\theta_i)LE(\beta_i)}\nonumber \\ =&  \prod_{p=-\frac{n-1}{2}}^{\frac{n-1}{2}}g^n_{p+\alpha}(r)\, \times  {}^n_{p,L}\bra {0} |\TT^\alpha (0) \Tilde{\TT}^\alpha (\ell) |{0}\ket_{p,L}^n
\end{align}
where we re-labelled the rapidities of the negatively charged intermediate states: $\beta_1 = \theta_{s+2},\dots, \beta_s = \theta_{2s+1}$. We can now make use of \eqref{eqn:g_function_identity} in the evaluation of the ratio, so that we finally obtain for the free fermionic one-particle states:
\beq
M^{1^+}_n(r;\alpha)=\prod_{p=-\frac{n-1}{2}}^{\frac{n-1}{2}} g_{p+\alpha}^n(r)= (1-r)^n + e^{2\pi i \alpha} r^n\,. 
\eeq
An analogous result can be obtained for a negatively charged particle, where the phase above picks up an extra minus sign. 

\subsection{Multi-Particle Excited States}
The anti-commuting nature of the creation/annihilation operators allows us to obtain an exact expression for the ratio of the charged moments in the fermionic case, which (unlike for the free boson) does not require a case-by-case calculation. This is because in the free fermion case, the structure of the states in the transformed base is extremely simple, as we shall see. 
We have non-vanishing two-point functions only with two kind of states:
\begin{itemize}
    \item $k$-particle excitations with distinct rapidities, irrespective of the charge signs;
    \item $2$-particle excitations with equal rapidities and different charge signs: $|1^+ 1^-\ket_L^n$\,.
\end{itemize}
Below, we consider in detail the case of $k$-particle states with distinct rapidities. Such states are written exactly as in the bosonic case:
\be
|1^{\epsilon_1} \, 1^{\epsilon_2} \dots 1^{\epsilon_k}\ket_L^n = \prod_{i=1}^k \prod_{j=1}^n (a_j^{\epsilon_i})^{\dagger} (\theta_i) |0\ket_L^n\,,
\ee
where $\epsilon_i = \pm 1$, $\theta_i \ne \theta_{i'}$ if $i\ne i'$. Unlike the bosonic case, however, all the operators anti-commute, so that we can make the ansatz: 
\be
\label{fermionic_multi-particle_state}
 |1^{\epsilon_1} \, 1^{\epsilon_2} \dots 1^{\epsilon_k}\ket_L^n=e^{i\kappa}\prod_{p=-\frac{n-1}{2}}^{\frac{n-1}{2}}\prod_{i=1}^k (\textfrak{a}_p^{\epsilon_i})^\dagger(\theta_i)|0\ket_{L}^n\,,
\ee
with the identification $(\textfrak{a}_p^+)^\dagger(\theta_i)=\textfrak{a}_p^\dagger(\theta_i)$, $(\textfrak{a}_p^-)^\dagger(\theta_i)=\textfrak{b}_p^\dagger(\theta_i)$ and the only unspecified parameter is the phase $\kappa = \kappa(k,n; \{\epsilon_i\})$. Notice that the order of the operators in the double product can be arbitrarily altered, resulting only in a change in the phase. Without giving a full proof of the validity of this formula, let us consider a few simple cases and introduce the notations $k^\pm$ to indicate the number of positively/negatively charged excitations in the state, with $k=k^+ + k^-$:
\begin{itemize}
    \item $n=2$, $k^+=2$:\\
    \begin{align}
    |1^+1^+\ket_L^2 &= \prod_{i=1}^{2}\prod_{j=1}^2\frac{1}{\sqrt{2}}\sum_{p=-\frac{1}{2}}^{\frac{1}{2}}e^{-\frac{2\pi i j p}{2}}\textfrak{a}_p^\dagger(\theta_i)|0\ket_{L}^2 \nonumber \\ 
    &= \frac{1}{4}\prod_{i=1}^2(i\textfrak{a}_{-\frac{1}{2}}^\dagger(\theta_i)-i\textfrak{a}_{\frac{1}{2}}^\dagger(\theta_i))(-\textfrak{a}_{-\frac{1}{2}}^\dagger(\theta_i)-\textfrak{a}_{\frac{1}{2}}^\dagger(\theta_i))|0\ket_{L}^2\nonumber \\
    &= \frac{1}{4}\prod_{i=1}^2(-2 i \textfrak{a}_{-\frac{1}{2}}^\dagger(\theta_i)\textfrak{a}_{\frac{1}{2}}^\dagger(\theta_i))|0\ket_{L}^2 = - \prod_{p=-\frac{1}{2}}^{\frac{1}{2}}\prod_{i=1}^2\textfrak{a}_p^\dagger(\theta_i)|0\ket_{L}^2 \nonumber
    \end{align}
    \item $n=2$, $k^+=2$, $k^-=1 $:\\
     \begin{align}
    |1^+1^+1^-\ket_L^2 &= \left(\prod_{i=1}^{2}\prod_{j=1}^2\frac{1}{\sqrt{2}}\sum_{p=-\frac{1}{2}}^{\frac{1}{2}}e^{-\frac{2\pi i j p}{2}}\textfrak{a}_p^\dagger(\theta_i)\right)\prod_{j=1}^2\frac{1}{\sqrt{2}}\sum_{p=-\frac{1}{2}}^{\frac{1}{2}}e^{\frac{2\pi i j p}{2}}\textfrak{b}_p^\dagger(\theta_3)|0\ket_{L}^2 \nonumber \\ 
    &=\left(\frac{1}{4}\prod_{i=1}^2(-2 i \textfrak{a}_{-\frac{1}{2}}^\dagger(\theta_i)\textfrak{a}_{\frac{1}{2}}^\dagger(\theta_i))\right)\frac{1}{2}(2i\textfrak{b}_{-\frac{1}{2}}^\dagger(\theta_3)\textfrak{b}_{\frac{1}{2}}^\dagger(\theta_3))|0\ket_{L}^2\nonumber \\&= -i\prod_{p=-\frac{1}{2}}^{\frac{1}{2}}\left(\prod_{i=1}^2\textfrak{a}_p^\dagger(\theta_i)\right)\textfrak{b}_p^\dagger(\theta_3)|0\ket_{L}^2\nonumber
    \end{align}
    \item $n=3$, $k^+=2$:\\
     \begin{align}
    |1^+1^+\ket_L^3 &= \prod_{i=1}^{2}\prod_{j=1}^3\frac{1}{\sqrt{3}}\sum_{p=-1}^1e^{-\frac{2\pi i j p}{3}}\textfrak{a}_p^\dagger(\theta_i)|0\ket_{L}^3 \nonumber \\ 
    &=\prod_{i=1}^2\frac{1}{3^\frac{3}{2}}\left(e^\frac{2\pi i}{3}\textfrak{a}_{-1}^\dagger(\theta_i)+\textfrak{a}_{0}^\dagger(\theta_i)+e^{-\frac{2\pi i}{3}}\textfrak{a}_{1}^\dagger(\theta_i)\right)\nonumber\\ &\times\left(e^\frac{4\pi i}{3}\textfrak{a}_{-1}^\dagger(\theta_i)+\textfrak{a}_{0}^\dagger(\theta_i)+e^{-\frac{4\pi i}{3}}\textfrak{a}_{1}^\dagger(\theta_i)\right) (\textfrak{a}_{-1}^\dagger(\theta_i)+\textfrak{a}_{0}^\dagger(\theta_i)+\textfrak{a}_{1}^\dagger(\theta_i))|0\ket_{L}^3 \nonumber \\
    &= - \prod_{p=-1}^1\prod_{i=1}^2\textfrak{a}_p^\dagger(\theta_i)|0\ket_{L}^3 \nonumber
    \end{align}
\end{itemize}
For a fixed value of $n$, the structure of more complicated states can be easily worked out following these simple examples. Using equation \eqref{fermionic_multi-particle_state} and the twist field factorisation \eqref{TTfactorizesector_fermion}, the two-point function reads:
\begin{align}
\label{multi-particle_two-point_function_1}
&_L^n\bra 1^{\epsilon_1} \, 1^{\epsilon_2} \dots 1^{\epsilon_k}|\TT^\alpha_n (0) \Tilde{\TT}^\alpha_n (\ell) |1^{\epsilon_1} \, 1^{\epsilon_2} \dots 1^{\epsilon_k}\ket_L^n \nonumber \\ = &\prod_{p=-\frac{n-1}{2}}^{\frac{n-1}{2}}{}^n_{p,L}\bra {0}| \textfrak{a}_{p} (\theta_1)  \dots \textfrak{a}_{p} (\theta_{k^+}) \textfrak{b}_{p} (\beta_1)  \dots \textfrak{b}_{p} (\beta_{k^-}) \TT_{p+\alpha} (0)  \TT_{-p-\alpha} (\ell)   \textfrak{a}_{p}^\dag (\theta_1) \dots \textfrak{a}_{p}^\dag (\theta_{k^+}) \textfrak{b}_{p}^\dag (\beta_1) \dots \textfrak{b}_{p}^\dag (\beta_{k^-})|{0}\ket_{p,L}^n \nonumber\\
= &\prod_{p=-\frac{n-1}{2}}^{\frac{n-1}{2}}\sum_{s=0}^{\infty}\sum_{\{J_i^\pm\}}\frac{1}{s!(s+q)!} \,e^{i\ell \left(\sum_{i=1}^{q+s}P(\tilde{\theta}_i) + \sum_{i=1}^s P(\tilde{\beta}_i) - \sum_{i=1}^{k^+} P(\theta_i) - \sum_{i=1}^{k^-} P(\beta_i) \right)} \nonumber \\
&\times {}^n_{p,L}\bra {0}| \textfrak{a}_{p} (\theta_1) \dots \textfrak{a}_{p} (\theta_k^+)\textfrak{b}_{p} (\beta_1)  \dots \textfrak{b}_{p} (\beta_{k^-}) \TT_{p+\alpha} (0) \textfrak{a}_p^\dag(\tilde{\theta}_1) \dots \textfrak{a}_p^\dag(\tilde{\theta}_{q+s})\textfrak{b}_p^\dag(\tilde{\beta}_1) \dots \textfrak{a}_p^\dag(\tilde{\beta}_s)|{0}\ket_{p,L}^n \nonumber\\ 
&\times {}^n_{p,L}\bra {0}| \textfrak{a}_{p} (\tilde{\theta}_1) \dots \textfrak{a}_{p}(\tilde{\theta}_{q+s})\textfrak{b}_{p} (\tilde{\beta}_1)  \dots \textfrak{b}_{p} (\tilde{\beta}_s) \TT_{-p-\alpha} (0) \textfrak{a}_p^\dag(\theta_1) \dots \textfrak{a}_p^\dag(\theta_{k^+})\textfrak{b}_p^\dag(\beta_1) \dots \textfrak{b}_p^\dag(\beta_{k^-})|{0}\ket_{p,L}^n \,.
\end{align}
In the expansion above we assumed the total charge of the excited state to be positive, $q\equiv k^+ - k^- > 0$. However, the computation steps are unchanged if one assumes $q < 0$, the only difference being in the structure of the resolution of the identity. Denoting $\hat{x} := x + i\pi$, the infinite-volume form factor corresponding to the first matrix element reads:
\be
F^{p+\alpha, n}_{q+2s+k}(\tilde{\theta}_1,\dots,\tilde{\theta}_{q+s}, \hat{\beta}_1,\dots,\hat{\beta}_{k^-};\tilde{\beta}_1,\dots, \tilde{\beta}_s, \hat{\theta}_1,\dots,\hat{\theta}_{k^+})\,,
\ee
where the total charge conservation is ensured by the equality $q+s+k^- = s+k^+$. When turning to the finite-volume, one needs to divide the previous infinite-volume form factor by a quantity:
\be
\left[\prod_{i=1}^{q+s} L E(\tilde{\theta}_i)\prod_{i=1}^{k^-} L E(\beta_i)\prod_{i=1}^{s} L E(\tilde{\beta}_i)\prod_{i=1}^{k^+}L E(\theta_i)\right]^\frac{1}{2}\,.
\ee
Taking into account also the conjugate form factor, this results into a factor $\sim L^{-q-2s-k}$ for every term in the expansion \eqref{multi-particle_two-point_function_1}, and the latter reads:
\begin{align}
\label{multi-particle_two-point_function_2}
 &\prod_{p=-\frac{n-1}{2}}^{\frac{n-1}{2}}\sum_{s=0}^{\infty}\sum_{\{J_i^\pm\}}\frac{1}{s!(s+q)!} \,e^{i\ell \left(\sum_{i=1}^{q+s}P(\tilde{\theta}_i) + \sum_{i=1}^s P(\tilde{\beta}_i) - \sum_{i=1}^{k^+} P(\theta_i) - \sum_{i=1}^{k^-} P(\beta_i) \right)} \nonumber \\
&\times \frac{\abs{F^{p+\alpha, n}_{q+2s+k}(\tilde{\theta}_1,\dots,\tilde{\theta}_{q+s}, \hat{\beta}_1,\dots,\hat{\beta}_{k^-};\tilde{\beta}_1,\dots, \tilde{\beta}_s, \hat{\theta}_1,\dots,\hat{\theta}_{k^+})}^2}{\prod_{i=1}^{q+s} L E(\tilde{\theta}_i)\prod_{i=1}^{k^-} L E(\beta_i)\prod_{i=1}^{s} L E(\tilde{\beta}_i)\prod_{i=1}^{k^+}L E(\theta_i)} \,.
\end{align}
In the $L \to \infty$ limit, the leading contributions are those coming from  simultaneous contractions in both form factors. In turn, in each form factor the simple poles arise from the pairings of the rapidities $\hat{\theta}_i$ with $\tilde{\theta}_i$ (these are at most $k^+$ contractions) and from the pairings of the rapidities $\hat{\beta}_i$ with $\tilde{\beta}_i$ (these are at most $k^-$ contractions). Again, these pairings have to be made simultaneously. The terms with $s < k^-$ (or equivalently $q+s < k^+$) do not contribute in the $L \to \infty$ limit, as they contain some extra factors of $L$ in the denominator. On the other hand, the terms with $s > k^-$ contain a sum of $k^+!k^-!$ products of the form:
\be
\abs{f^n_{p+\alpha}(\tilde{\theta}_i - \hat{\theta}_j)f^n_{p+\alpha}(\hat{\beta}_{i'}-\tilde{\beta}_{j'})}^2\times \text{\,residual form factors}  \,. \nonumber
\ee
By making use of \eqref{fermionic_two-particle_ff_simple_pole}, the simultaneous expansion around the poles leads to the following leading contribution:
\begin{align}
&\sum_{J_i^+, J_{j'}^-}\frac{\abs{f^n_{p+\alpha}(\tilde{\theta}_i - \hat{\theta}_j)f^n_{p+\alpha}(\hat{\beta}_{i'}-\tilde{\beta}_{j'})}^2}{\cosh{\tilde{\theta}_i}\cosh{\theta_i}\cosh{\beta_{i'}}\cosh{\tilde{\beta}_{j'}}}\,  e^{i\ell[P(\tilde{\theta}_i)-P(\theta_j)+ P(\tilde{\beta}_{j'})-P(\beta_{i'})]} \nonumber \\ 
& \sim (m L)^4 \sum_{J_i^+ \in \mathbb{Z}} \frac{\sin^2{\frac{\pi(p+\alpha)}{n}}}{\pi^2}\frac{e^{2\pi i r (J_i^+ - I_j + \frac{p+\alpha}{n})}}{(J_i^+ - I_j + \frac{p+\alpha}{n})^2} 
\sum_{J_{j'}^- \in \mathbb{Z}} \frac{\sin^2{\frac{\pi(p+\alpha)}{n}}}{\pi^2}\frac{e^{2\pi i r (J_{j'}^- - I_{i'} - \frac{p+\alpha}{n})}}{(J_{j'}^- - I_{i'} - \frac{p+\alpha}{n})^2} \nonumber \\&= (m L)^4 g^n_{p+\alpha}(r)g^n_{-p - \alpha}(r)\,.
\end{align}
Since each of the $k^+!k^-!$ terms in the sum contains exactly $k^+$ functions $g^n_{p+\alpha}$ and $k^-$ functions $g^n_{-p-\alpha}$ we have, relabelling $m = s-k^-$ and in the limit $L\,,\ell \to \infty$, $r$ fixed:
\beqa
\label{multi-particle_two-point_function_fermionic-final}
&& _L^n\bra 1^{\epsilon_1} \, 1^{\epsilon_2} \dots 1^{\epsilon_k} |\TT^\alpha_n (0) \Tilde{\TT}^\alpha_n (\ell) |1^{\epsilon_1} \, 1^{\epsilon_2} \dots 1^{\epsilon_k}\ket_L^n \nonumber \\ && =\prod_{p=-\frac{n-1}{2}}^{\frac{n-1}{2}}\sum_{m=0}^{\infty}\sum_{\{J_i^\pm\}}\frac{1}{(m!)^2} (g_{p+\alpha}^n(r))^{k^+} (g_{-p-\alpha}^n(r))^{k^-} \nonumber\\ 
&&\times e^{i\ell \sum_{i=1}^m(P(\tilde{\theta}_i) + P(\tilde{\beta}_i))} \, \frac{\abs{F^{p+\alpha, n}_{2m}(\tilde{\theta}_1,\dots,\tilde{\theta}_m;\tilde{\beta}_1,\dots,\tilde{\beta}_m)}^2}{\prod_{i=1}^m LE(\tilde{\theta}_i)LE(\tilde{\beta}_i)} \nonumber \\
&&=\left(\prod_{p=-\frac{n-1}{2}}^{\frac{n-1}{2}}g_{p+\alpha}^n(r)\right)^{k^+}\left(\prod_{p=-\frac{n-1}{2}}^{\frac{n-1}{2}}g_{-p-\alpha}^n(r)\right)^{k^-}\, _L^n\bra 0 |\TT^\alpha_n (0) \Tilde{\TT}^\alpha_n (\ell) |0\ket_L^n
\eeqa 
and therefore, thanks to \eqref{TTfactorizesector_fermion}, the ratio of charged moments is:
\beq
M^{1^\pm1^\pm\dots 1^\pm}_n(r;\alpha)= ((1-r)^n + e^{2\pi i \alpha} r^n)^{k^+}((1-r)^n + e^{-2\pi i \alpha} r^n)^{k^-}\,. 
\eeq

\section{Finite-Volume Two-Point Function in the Ground State}
\label{finiteVexp}
In this Appendix we want to investigate the large volume expansion of the correlator
\beq
\label{eqn:finite_volume_1}
{}_L^n\bra 0|\TT_n^\alpha(0)\tilde{\TT}_n^\alpha(rL)|0\ket_L^n\,,
\eeq
which is the denominator of the ratio of charged moments $M_n^\Psi(r;\alpha)$ in \eqref{eqn:moments_of_ratio}. We show that, as expected, the leading contribution as $L \to + \infty$ is given by the squared modulus of the vacuum expectation value of the composite twist field $\TT^\alpha_n$, and we compute the first finite-volume corrections to this quantity. The calculations are carried out in the fermionic case, but they apply to the free boson case with few changes.

The first step in the evaluation of \eqref{eqn:finite_volume_1} is as usual the insertion of a projection onto asymptotic states:
\begin{align}
\label{eqn:finite_volume_2}
_L^n\bra {0} |\TT^\alpha_n (0) \Tilde{\TT}^\alpha_n (\ell) |{0}\ket_L^n =& \prod_{p=-\frac{n-1}{2}}^{\frac{n-1}{2}}{}^n_{p,L}\bra {0}|   \TT_{p+\alpha} (0) \; \TT_{-p-\alpha} (\ell)  \;   |{0}\ket_{p,L}^n \nonumber\\
= &\prod_{p=-\frac{n-1}{2}}^{\frac{n-1}{2}}\sum_{s=0}^{\infty}\sum_{\{J_i^\pm\}}\frac{1}{(s!)^2}{}^n_{p,L}\bra {0}|  \TT_{p+\alpha} (0) \; \textfrak{a}_{p}^\dag (\theta_1)\dots \textfrak{a}_{p}^\dag (\theta_s)\textfrak{b}_{p}^\dag (\beta_1)\dots\textfrak{b}_{p}^\dag (\beta_s)|{0}\ket_{p,L}^n \nonumber\\ &\times {}^n_{p,L}\bra {0}| \textfrak{a}_{p} \; (\theta_1)  \;\dots \textfrak{a}_{p} (\theta_s)\textfrak{b}_{p} (\beta_1)\dots\textfrak{b}_{p} (\beta_s) \TT_{-p-\alpha} (0) |{0}\ket_{p,L}^n e^{i\ell \sum_{i=1}^{s}(P(\theta_i)+P(\beta_i))} \nonumber \\
 =&\prod_{p=-\frac{n-1}{2}}^{\frac{n-1}{2}}\sum_{s=0}^{\infty}\sum_{\{J_i^\pm\}} \frac{\abs{{F^{p+\alpha,n}_{2s}(\theta_1,\dots,\theta_s;\beta_1,\dots,\beta_s)}}^2}{(s!)^2\prod_{i=1}^{s}L^2E(\theta_i)E(\beta_i)}e^{i\ell \sum_{i=1}^{s}(P(\theta_i)+P(\beta_i))}\,.
\end{align}
Notice that the equal number of particles of the two types is dictated by the fact that twist fields preserve the total charge of the state. In a free fermion theory, the infinite-volume form factor of an even number of particle is given by Wick's theorem:
\beq
F^{p+\alpha,n}_{2s}(\theta_1,\dots,\theta_s;\beta_1,\dots,\beta_s) = \sum_{\sigma \in \mathcal{P}_s} \tau_{p+\alpha}\, \sgn{\sigma}  \prod_{i=1}^s \frac{f^n_{p+\alpha}(\theta_{\sigma(i)}-\beta_i)}{\tau_{p+\alpha}}
\eeq
where the fermionic two-particle form factor was given in (\ref{FFfer}) so that the squared modulus of the $2s$-particle form factor is a sum of $(s!)^2$ terms, each of which is a product of $s$ terms of the type
\beq
\frac{f_{p+\alpha}^n(\theta_i - \beta_j)f_{p+\alpha}^n(\theta_k - \beta_j)^*}{\abs{\tau_{p+\alpha}}^2}=  \sin^2\frac{\pi(p+\alpha)}{n}\frac{e^{\frac{p+\alpha}{n}(\theta_i + \theta_k - 2\beta_j)}}{\cosh\frac{\theta_i - \beta_j}{2}\cosh\frac{\theta_k - \beta_j}{2}}\,.
\eeq
We can therefore explicitly rewrite \eqref{eqn:finite_volume_2} as:
\begin{align}
\label{eqn:finite_volume_3}
_L^n\bra {0} |\TT^\alpha_n (0) \Tilde{\TT}^\alpha_n (\ell) |{0}\ket_L^n &= \prod_{p=-\frac{n-1}{2}}^{\frac{n-1}{2}}\abs{\tau_{p+\alpha}}^2\sum_{s=0}^{\infty}\frac{\sin^{2s}\frac{\pi(p+\alpha)}{n}}{(s!)^2(mL)^{2s}}\sum_{\{J_i^\pm\}}\sum_{\sigma , \omega \in \mathcal{P}_s} \sgn{\sigma} \sgn{\omega}
\nonumber\\ &\times \prod_{i=1}^s\frac{e^{i\ell(P(\theta_i)+P(\beta_i))}}{\cosh{\theta_i}\cosh{\beta_i}}\frac{e^{\frac{p+\alpha}{n}(\theta_{\sigma(i)} + \theta_{\omega(i)} - 2\beta_i)}}{\cosh\frac{\theta_{\sigma(i)} - \beta_i}{2}\cosh\frac{\theta_{\omega(i)} - \beta_i}{2}}\,.
\end{align}
From this expression we easily see that the vacuum expectation value is corrected by contributions of multi-particle states, and that in general every $2s$-particle state (containing $s$ particles with positive charge and $s$ particles with negative charge) contributes with a leading large-volume term $\sim (mL)^{-2s}$. Further corrections can be obtained by working out how the product in the second line of \eqref{eqn:finite_volume_3} depends on $L$. This is done by solving the Bethe equations for $\sinh{\theta_i}$ and $\sinh{\beta_i}$:
\beq
\sinh{\theta_i} = \frac{2\pi\left(J_i^+ + \frac{p+\alpha}{n}\right)}{mL} \equiv \frac{c_i^+}{mL}\,,
\eeq
\beq
\sinh{\beta_i} = \frac{2\pi\left(J_i^- - \frac{p+\alpha}{n}\right)}{mL} \equiv \frac{c_i^-}{mL}\,.
\eeq
Where either $J_i^+\,,J_i^- \in \mathbb{Z}$ or $J_i^+\,,J_i^- \in \mathbb{Z}+\frac{1}{2}$. Let us consider in detail the expansion up to $s=1$ terms, assuming that $J_1^+\,,J_1^- \in \mathbb{Z}$. Some elementary algebra shows that:
\beq
\frac{e^{2(\theta_1 - \beta_1)\frac{p+\alpha}{n}}}{\cosh^2{\frac{\theta_1-\beta_1}{2}}}=1+ 2\left(\frac{p+\alpha}{n}\right)\frac{c_1^+-c_1^-}{mL} + \mathcal{O}\left(\frac{1}{(mL)^2}\right)
\eeq
and:
\beq
\frac{e^{i\ell(P(\theta_1)+P(\beta_1))}}{\cosh{\theta_1}\cosh{\beta_1}}=e^{2\pi i r(J_1^++J_1^-)}\left(1-\frac{(c_1^+)^2+(c_1^-)^2}{2(mL)^2}+\mathcal{O}\left(\frac{1}{(mL)^4}\right)\right)\,.
\eeq
Thus we have:
\begin{align}
\label{eqn:finite_volume_4}
&_L^n\bra {0} |\TT^\alpha_n (0) \Tilde{\TT}^\alpha_n (\ell) |{0}\ket_L^n =\nonumber \\ &\prod_{p=-\frac{n-1}{2}}^{\frac{n-1}{2}}\abs{\tau_{p+\alpha}}^2\left[1 + \frac{\sin^2\frac{\pi(p+\alpha)}{n}}{(mL)^2} \sum_{J_1^+,J_1^- \in \mathbb{Z}}e^{2\pi i r(J_1^+ + J_1^-)}\left(1+ 2\left(\frac{p+\alpha}{n}\right)\frac{c_1^+-c_1^-}{mL} + \mathcal{O}\left(\frac{1}{(mL)^2}\right)\right)\right]\,.
\end{align}
We immediately notice that there is no contribution of order $\sim (mL)^{-2}$, as we can regularise the non convergent double sum:
\beq
\sum_{J_1^+,J_1^- \in \mathbb{Z}}e^{2\pi i r(J_1^+ + J_1^-)} = \left(\sum_{J \in \mathbb{Z}}e^{2\pi i r J}\right)^2=0\,,
\eeq
as was shown in Appendix C of \cite{excited2}. 
Considering the $\sim (mL)^{-3}$ term we have to evaluate the double sum:
\begin{equation*}
\sum_{J_1^+,J_1^- \in \mathbb{Z}}e^{2\pi i r(J_1^+ + J_1^-)}(J_1^+-J_1^-)\,,
\end{equation*}
which can also be shown to be zero. This means that first finite volume correction to the ratio of moments of the SREE 
is of order $(mL)^{-4}$, which is a bit more involved as it picks a contribution also from the first term with $s=2$. Investigating these corrections is beyond the scope of this paper, but we expect to return to this problem and consider finite volume corrections to the excess (total) entropy in future work. 


\end{document}